\documentclass[preprint2]{aastex}

\shorttitle{Paper I}
\shortauthors{Martin}

\begin{document}


\title{The Origins and Evolutionary Status of B Stars Found Far From the Galactic Plane I:  Composition and Spectral Features\altaffilmark{2}}


\author{J. C. Martin\altaffilmark{1}}
\affil{Case Western Reserve University\\Cleveland, OH 44106}
\email{jmartin@astro.umn.edu}

\altaffiltext{1}{Currently at the University of Minnesota, Minneapolis, MN 55455.}
\altaffiltext{2}{Based on observations made at
 the 2.1-m Otto Struve Telescope of McDonald Observatory operated by the University of Texas at Austin}



\begin{abstract}
The existence of faint blue stars far above the galactic plane, which have spectra that are similar to nearby Population I B stars, present several interesting questions.  Among them are:  Can a Population I B star travel from the disk to a position many kiloparsecs above the plane in a relatively short main sequence lifetime?  Is it possible that single massive star formation is occurring far from the galactic plane?  Are these objects something else masquerading as main sequence B stars?  This paper (the first of two) analyzes the abundances of a sample of these stars and reveals several which are chemically similar to nearby Population I B stars while others clearly have abundance patterns more like those expected in BHB or PAGB stars.  Several of those with old evolved star abundances also have interesting features of note in their spectra.  We also consider why this sample does not have any classical Be stars and identify at least two nearby solar metallicity BHB stars.

\end{abstract}




\section{Introduction}
\subsection{History}
Faint blue stars at high galactic latitudes were first discovered in a survey 
conducted by \citet{humason47}.  Soon after, \citet{bidelman48} brought 
attention to a number of stars of early spectral type with heliocentric velocities greater than 100 km/s.  Their faint apparent magnitudes (sizable distances) and high velocities led to the conclusion that these objects, which appear at low spectroscopic dispersions to be Population I stars, are in fact a part of the galactic halo. 

There are currently three basic explanations for B stars found far from the galactic 
plane:
\begin{enumerate}
\item{``Normal'' massive Population I stars ejected from the galactic disk, }
\item{Older evolved stars (i.e. blue horizontal branch (BHB) or post 
asymptotic giant branch (PAGB) stars), or} 
\item{Young massive stars formed in situ in the galactic halo. }
\end{enumerate}
We will now briefly discuss the history of these scenarios and characteristics that can be used to distinguish which stars have been affected by each scenario.

\citet{gs74} studied a sample of faint blue stars within 30 degrees of the 
galactic poles and found that 50 of the 189 objects surveyed at low spectroscopic dispersion were indistinguishable from normal Population I B stars.  The radial velocities of these stars implied that twenty of them took longer than their projected main sequence lifetimes to travel from the disk to their present location.  Since in situ star formation in the galactic halo seemed an impossible scenario to them, \citet{gs74} concluded that several of these seemingly normal Population I main sequence stars are in fact low mass evolved stars.

Subsequent photometric and spectroscopic surveys continued to identify significant numbers of high galactic latitude B stars as "normal" Population  I stars until \citet{tobin87} showed that many of these studies could have mistaken older evolved stars for young main sequence stars.  Armed with this information, researchers interested in  post asymptotic giant branch (PAGB) stars saw this population of B stars far above the galactic plane as a potential boon to the study of rare and poorly defined stages of stellar evolution \citep{mccausland92}. The possibility that many of these stars could be old evolved stars masquerading as ``normal'' Population I stars is the impetus for this research and other studies including \citet{magee01}, \citet{ramspeck01a}, and \citet{ramspeck01b}.  These new studies, realize the importance of using many different analysis tools in concert since no one indicator alone has proven able to characterize the B stars found far from the galactic plane.

The observed spectra of stars are largely a function of effective temperature 
and surface gravity. Unfortunately, these parameters alone are not enough to 
distinguish high mass hydrogen burning main sequence stars from low mass post 
asymptotic giant branch (PAGB) stars or low mass helium burning blue 
horizontal branch (BHB) stars.  Hot PAGB stars may traverse one of many 
evolutionary tracks with effective temperatures and surface gravities that are 
comparable to stars on the main sequence.   BHB stars with effective 
temperatures between 10,000 K and 20,000 K, where the horizontal branch 
crosses the main sequence, are also easily confused with main sequence B stars 
\citep{tobin87}.  In this regime, stars that otherwise have the same 
atmospheric parameters can have large differences in mass and luminosity 
depending on their evolutionary status.  Temperatures and gravities can be 
determined for these stars by various means (photometry, absolute flux 
fitting, fitting of spectral features) but their masses, distances, and
intrinsic luminosities remain in most cases only an educated guess.  Fortunately, their photospheric compositions and full space velocities provide additional clues the origin and evolutionary stage of a star.

Runaway Population I stars ejected from the galactic disk should have 
photospheric elemental abundances which are, for the most part, 
indistinguishable from their parent population (nearby Population I B stars in 
the galactic disk).  At the same time, they should have space velocities which allow them to traverse large distances from the galactic plane to their present location within their main sequence lifetime.  

Older evolved stars should exhibit photospheric abundances patterns which are markedly different from Population I stars due to their initial composition and processes which occurred earlier in their lifetimes such as mixing and dust grain formation.  Many older evolved stars are also less luminous than main sequence stars with the same effective temperature. Overestimating their luminosity and photometric distance by assuming that they are main sequence stars will inflate their space velocities causing them to stand out kinematically from other populations.

The final possibility is that these are massive stars which have formed far out of the galactic disk.  Massive stars formed in situ in the halo should have photospheric elemental abundances which are consistent with the composition of gas 
clouds found far from the galactic plane (i.e. Intermediate Velocity Clouds (IVCs) or High Velocity Clouds (HVCs)).  These stars should kinematically belong to the halo population and need not have velocities which allow them to have traveled from the disk to their present location within their main sequence lifetime.

We undertook a multifaceted study which is designed to try to unravel the various possibilities for these stars.  Due to the extent of the work, it has been divided into two papers for publication.  
\begin{enumerate}
\item{This paper analyzes the photospheric elemental abundances and other spectral characteristic of a sample of high galactic latitude B stars.}
\item{The second paper will analyze the kinematics of the stars in the sample and then use a combination that data and the data from this paper to classify the origin and nature of each star in the sample.}
\end{enumerate}

\subsection{Distinguishing Stellar Populations Using Relative Photospheric Abundance Analysis}

In order to compare the abundances of stars they must be subjected to the same analysis because absolute abundances are subject to systematic errors which are difficult to quantify.  This means that a control sample must be analyzed if one wants to make a valid comparison between the stars in a sample and another stellar population. Many past studies of high latitude B stars neglected to use a control sample in their analysis and as such were unable to make strong conclusions about the chemical composition of those stars relative to nearby B stars.  In contrast, more recent work by \citet{hambly97}, \citet{rolleston99}, and \citet{magee01} have compared the abundances of B stars found far from the galactic plane with control samples of nearby Population I B stars.  These studies have shown that many of the high galactic latitude B stars have abundances which are strikingly similar to nearby Population I B stars, with some small indication of slight deficits or enhancements in a few elements.  

BHB stars have average rotational velocities which are significantly lower 
than main sequence stars with the same effective temperatures 
\citep{peterson93,peterson95}.  \citet{peterson83} showed that stars on the 
horizontal branch in globular clusters have projected rotational velocities 
(vsin(i)) smaller than 40 km/s, as opposed to main sequence B stars which have 
average rotational velocities on the order of 100 km/s \citep{abt02}.  While 
fast rotating BHB stars appear to be ruled out, main sequence B stars with 
small vsin(i) are not (for example HR 6165, $\gamma$ Peg, HR 8768).   For the time 
being, we will ignore vsin(i) as a selection criterion so that the final selection of runaway B stars is unbiased with respect to this parameter.  Detailed abundance analysis should be able to differentiate between these populations since conventional wisdom holds that the abundance pattern of elements in BHB stars should be significantly different from main sequence stars due to their advanced evolutionary state. 

The abundances of evolved stars differ from those of main sequence B stars for 
a variety of reasons including: 
\begin{itemize}
\item{Initial composition,}
\item{Enhancement or depletion of specific elements by nuclear processing, and}
\item{Depletion of refractory elements by dust grain formation. }
\end{itemize}
It is not uncommon to find stars in similar advanced evolutionary states which have differing abundance patterns due to variation in the processes which combined to create the composition of their stellar atmospheres.  For the most part, the atmospheres of main sequence stars are influenced by their initial composition, although mixing and other factors may have a small effect.  Still there is relatively little variation in the abundances of B stars recently formed in the solar neighborhood between 6 kpc and 10 kpc from the galactic center where the disk abundance gradient appears to be almost flat \citep{kilian92,gl92,daflon03,andrievsky04}.  It is reasonable to assume that most runaways which we observe at high galactic latitudes come from this region since most of their ejection velocity is probably directed perpendicular to the galactic plane keeping them close to the galactocentric radius which they formed at.  Therefore, under ideal conditions abundance anomalies are a certain way to separate high mass main sequence stars from low mass evolved stars.

There is the unexcluded possibility that some high latitude B stars are the product of ongoing in situ star formation in the galactic halo.  
\citet{vanwoerden93} and \citet{christodoulou97} report that the conditions 
thought to exist in intermediate and high velocity clouds (IVCs  and HVCs) 
found high above the galactic plane are conducive to the formation of 
individual massive stars.  Even if the conditions exist in IVCs and HVCs for 
ongoing star formation, there is no observational evidence of star formation 
in these clouds \citep{hopp02,willman02,ivezic97,hambly96}.  Star formation 
apparently is occurring under what are thought to be similar environmental 
conditions in the H I bridge between the LMC and SMC \citep{christodoulou97} 
and \citet{comeron01} report signs of star formation occurring in the halo 
of the edge-on spiral galaxy NGC 253.  However, it should be noted that NGC 
253 is a starburst galaxy so those results may not be applicable to the Milky 
Way.  \citet{wakker01} found that HVCs and IVCs have significantly metal poor 
chemical compositions.  Therefore, any star formed in the halo would have to have  photospheric abundances which are comparable to these clouds.  As a result, we expect that B stars formed in situ in the halo should be metal poor by a factor of ten to one hundred with respect to B stars formed in the disk.

\section{The Sample \& Data}

Forty nine (49) high galactic latitude B stars were identified in the 
Hipparcos catalog \citep{hip} for this study (Table \ref{tab1}).  The sample is 
restricted to stars above declination -30 degrees which can be observed with 
the McDonald Observatory 2.1-m Struve Reflector using the Sandiford Echelle 
Spectrograph \citep{sandiford}.  Thirty two (32) are from the 
\citet{gs74} survey list of ``normal'' B stars.  The other seventeen (17) 
stars are B stars from the Hipparcos Catalog \citep{hip} which have 
spectroscopic parallaxes and galactic latitudes placing them at least one 
kiloparsec above the galactic plane.  Contamination of the sample by nearby 
subdwarfs was minimized by rejecting any star not classified by \citet{gs74} 
which has a significant Hipparcos geometric parallax ($\pi > 2\sigma_{\pi}$).  
Some of the stars in the sample are not strictly at what some may consider to 
be high galactic latitude, but all of them are suspected to be a significant 
distance above the galactic plane.  

In addition to the sample of study, three control samples were designated to 
serve as comparisons for the abundance analysis (Table \ref{tab2}).  An abundance control sample of nearby Population I B stars was taken from the Yale Bright Star Catalog \citep{bsc}.  Small vsin(i) was favored in the nearby Population I sample because weak metal lines can be measured to higher accuracy in stars 
with smaller rotational broadening.  A few more rapid rotators were included 
as well because a number of the stars in the sample of study have significant 
vsin(i) values.  In most cases, stellar rotation has no significant effect on the abundance analysis.  However, the most rapid rotators were excluded because at the target signal to noise (100:1) it is exceedingly difficult to measure lines in stars rotating in excess of 150 km/s.  Two additional 
samples were selected to serve as comparisons in the abundance analysis: a 
sample of blue horizontal branch (BHB) stars was selected from the brighter 
objects listed in \citet{gs74} and a sample of post asymptotic giant branch (PAGB) stars of spectral type B9 or hotter was taken from a list provided by \citet{luckpagb}.

\subsection{Observations \& Reductions}

High resolution spectra (R=60000) were obtained for all the stars in the 
control samples and forty three (43) of the stars in the main sample on six 
separate observing runs using the McDonald Observatory 2.1-m 
Struve Reflector with the Sandiford Echelle Spectrograph 
\citep{sandiford}.  Constraints on observing time meant that no spectra 
were observed for six stars in the sample (HD 8323, BD +16 2114, BD +49 2137, 
HD 121968, HD 125924, and BD +20 3004).  The abundances of three of the stars observed were not analyzed because of obstacles that prevented their spectra from being measured:  Feige 23 (had no measurable metal lines), BD +13 3224 (is a short period pulsating helium dwarf V652 HER), AG +03 2773 (is not a B star, see section 3.1).  Even though these stars are excluded from the abundance analysis, they are included in other aspects of this study.

All the spectra were reduced and extracted using IRAF 
\footnote{IRAF is distributed by the National Optical Astronomy Observatories, which are operated by the Association of Universities for Research in Astronomy, Inc., under cooperative agreement with the National Science Foundation.}.  
The extracted spectra were continuum normalized, wavelength calibrated, and 
had equivalent widths and feature depths determined using the ASP software package of \citet{luckasp}.  Some of the spectra were binned in order to increase the signal to noise of the data.  After combining exposures and binning the data almost all of the spectra have 
a signal to noise of at least 100:1. 

The equivalent widths were measured interactively by direct integration of 
the line profiles in the spectra.  Only single unblended lines and lightly blended 
lines which could be de-blended were measured.  The wavelength overlap of the 
Echelle orders makes it possible to measure the equivalent width of some lines 
on two or more adjacent orders.  Typically, multiple measures of the same feature varied on the order of the noise level (see below).  Individual measures of the equivalent width from adjacent orders and multiple observations were averaged to improve accuracy.  Lines with observed equivalent widths which are smaller than the noise level in the data have been dropped from the analysis.  The minimum measurable equivalent width is calculated from the noise contribution to the integrated line profile as the area of a Gaussian with the depth of the noise level and the FWHM of the lines in the spectra.  Expressing the FWHM as a function of line width in velocity space (vsin(i)) and normalizing the signal to one yields:\\
\begin{equation}
\sigma_{EW}=\frac{3 \lambda_{0}}{2(S/N)}\times \frac{vsin(i)}{c}
\end{equation}
Where $\lambda_{0}$ is the wavelength of the line center and c is the speed of 
light.   Therefore, a spectrum with the target S/N of 100 at 4500 \mbox{\AA} has 2.3 m\mbox{\AA} of noise for each 10 km/s of line broadening.  This means that typically the noise level ranges from a few milliangstroms for the spectrum of a narrow lined star up to 50 m\mbox{\AA} or more for a spectrum with low S/N and very broad lines.  

For quality control, the equivalent widths measured from spectra of HR 6165 by this study were compared with those measured by \citet{hambly97} and \citet{kilian89}.  There is a negligible average difference ($0.3 \pm 4.5$ m\mbox{\AA}) between the equivalent widths of the 93 lines measured both this study and \citet{hambly97}.  The small systematic difference ($4.6 \pm 4.8$ m\mbox{\AA}) between the equivalent widths measured by both this study and \citet{kilian89} is most likely attributable to a difference in spectral resolution and subjective uncertainty associated with continuum setting.  In any case, 4.6 m\mbox{\AA} is not significant compared to the noise level encountered in most of the data.

\section{Stellar Atmospheric Parameters}

\subsection{Effective Temperature \& Surface Gravity}

The effective temperature and surface gravity of the stars in the samples were 
determined by combining results obtained from a number of different 
photometric indices and flux fitting methods.  The photometry is taken from 
GCPD\footnote{http://obswww.unige.ch/gcpd/gcpd.html} database 
\citep{mermilliod97} and the references cited therein.  The photometry is 
corrected for interstellar extinction using the E(B-V) values listed in Table \ref{tab3} (references are listed in the table) and extended to other wavelengths and colors using excess ratios derived from the interstellar extinction curve of \citet{fitzpatrick99} (assuming $\frac{E(B-V)}{A_V}=3.1$).  

The results of the photometric temperature and gravity relations are presented in Table \ref{tab3}.  The Johnson (B-V) color-temperature relation \citep{nsw93} should be considered only a crude estimate because the large band-passes and effective wavelengths of the Johnson UBV filters make them poorly suited to measure blackbody curves of B stars.  The Str\"{o}mgren temperature relation of \citet{nsw93} was selected over other Str\"{o}mgren methods because, unlike other relations, it is independent of the $H\beta$ index which can be influenced unpredictably by Balmer line emission which is common in B stars. The Geneva Photometry for each star was processed using the temperature and surface gravity relation of \citet{kunzli97}.

Surface gravities were also measured from the Balmer line profiles by comparing the observed spectra to synthetic profiles generated using ATLAS9 and BALMER9 \citep{atlas} assuming the effective temperatures derived from the photometry.  The wings of the Balmer lines cover most of the Echelle orders so the continua are set by rescaling the continua from the previous or subsequent orders without a hydrogen line.  The estimated accuracy of this method is $\pm0.25$ dex or better.

\begin{figure}
\figurenum{1}
\label{fig1}
\includegraphics[angle=90,scale=0.3]{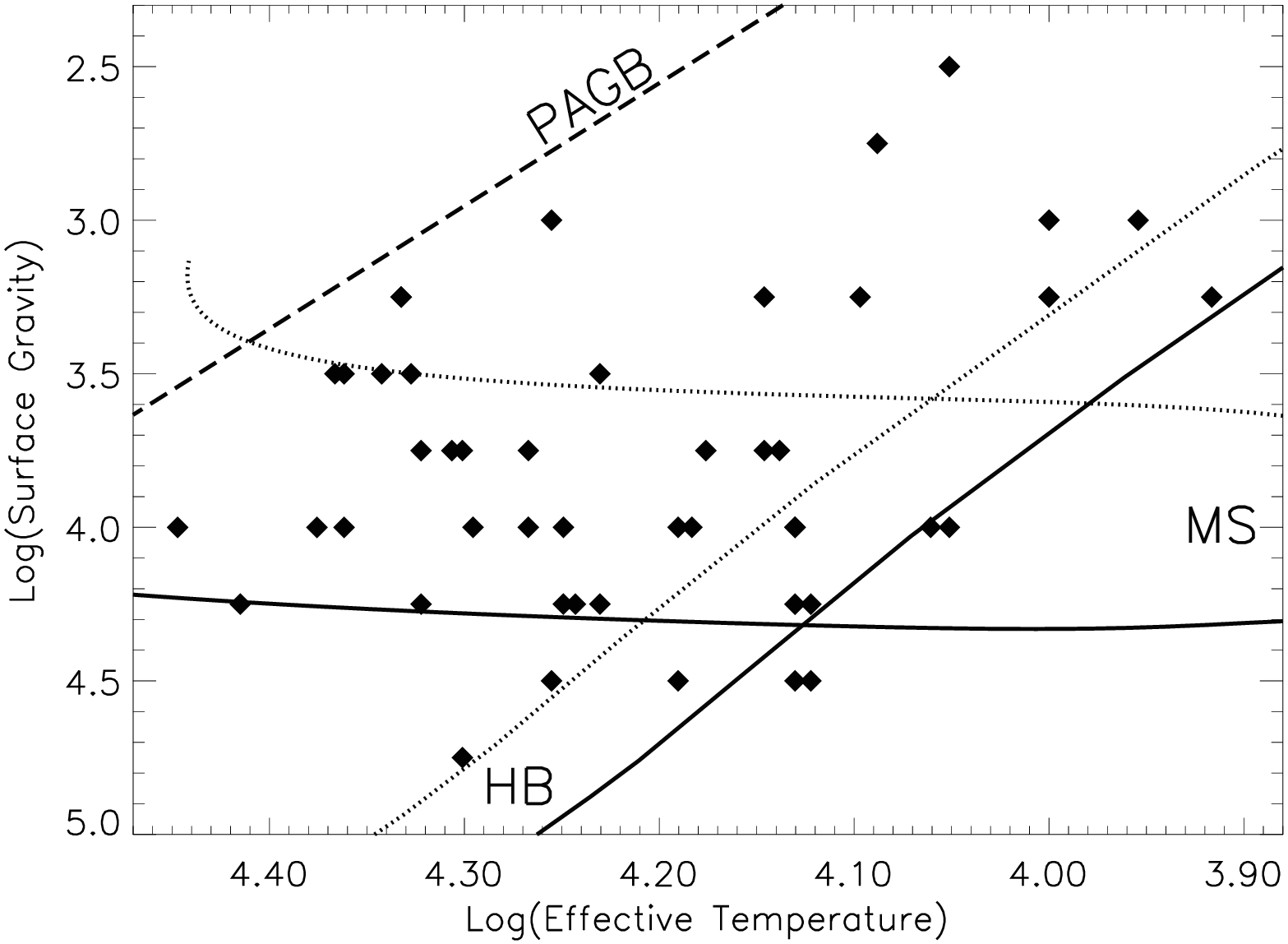}
\caption{A plot of the effective temperatures and surface gravities of the program stars (diamonds) compared to the position of the zero age (solid line) and terminal age (dotted line) Main Sequence \citep{zams,tams}, the zero age (solid line) and terminal age (dotted line) Horizontal Branch \citep{dorman93}, and a representative post-asymptotic branch track for a 1.5 M$_{\sun}$ star (dashed line) \citep{pagbtrack}.  All the tracks represented are for solar metallicity ([Fe/H] = 0).}
\end{figure}

The results from the individual temperature and gravity relations were combined into single values. The greatest weight given to the effective temperatures determined using Geneva photometry with the Str\"{o}mgren temperature estimate being a close second.  For surface gravities, the highest weights were given to values derived from Balmer line fitting and/or Geneva photometry.  Most importantly, the temperature and gravity of each star (Table \ref{tab4}) were accepted only if the observed spectrum could be reasonably synthesized by MOOG (see section 4.2) using an atmospheric model with those parameters.  It should be noted that the spectra of stars with significant rotational broadening are reasonably fit by a wider range of atmospheric parameters than stars with sharp spectral features.  Therefore, there is less certainty in the atmospheric parameters of more rapidly rotating stars.  Figure \ref{fig1} shows the effective temperature and surface gravity of each star plotted in relation to the main sequence and horizontal branch.

Unfortunately BD +61 996, HIP 41979 and AG +03 2773 have only (B-V) photometry and the temperatures derived from that data do not produce synthetic spectra which match the observed spectra.  For HIP 41979 and BD +61 996, the temperatures from the (B-V) data and the gravities from the Balmer profiles were used as starting points to determine the effective temperatures by matching the features in the observed spectra to a series of synthetic spectra produced by MOOG.  AG +03 2773 has a unique spectrum with many strong rotationally broadened absorption lines making it nearly impossible to set the continuum.  Furthermore, the (B-V) color (0.275) is not consistent with a star of its spectral type (B5) and interstellar extinction along the line of sight is insufficient to explain this discrepancy.  The most likely explanation is that this star is actually a late type A or early type F star which has been misclassified.

NLTE effects may impact the photometric effective temperature and surface 
gravity relations since most of them rely on theoretical colors calculated 
from stellar atmospheres which assume LTE.  The role of NLTE effects with 
respect to these calibrations is more significant in hotter ($\geq$ 20000 K) 
and/or lower gravity stars.  \citet{kilian91a} found that temperature 
estimates for B stars made from the Str\"{o}mgren [c1] index are 
systematically 2000 K cooler than temperature estimates made from \ion{H}{1} and \ion{Si}{3} lines after accounting for NLTE effects.  However, \citet{vrancken96} found that \citet{kilian91a} had overestimated this difference by 
at least 300 K to 500 K.  Fortunately, \citet{sigut96} found no similar offset between NLTE temperature estimates and the \citet{nsw93} [u-b] temperature 
relation employed in this study.  The exact effect of NLTE on derived stellar parameters is an unresolved subject of ongoing study.  The best course of action is to be conscious of the fact that, particularly in hot super giants, there may be a tendency to overestimate the stellar effective temperature.

\subsection{Microturbulence}

The microturbulent velocity ($\xi_{t}$) is determined by finding the value at 
which the slope of abundance versus equivalent width for the lines of 
a given atomic species is zero.  Species with many lines that span a large range of equivalent widths are the best barometers of microturbulence.  For hotter B Stars, \ion{O}{2}, \ion{N}{2}, and \ion{Si}{3} have well populated curves of growth that are ideal for determining the microturbulence.  For cooler B stars, \ion{Fe}{2}, \ion{Cr}{2}, and \ion{Ti}{2} are good species to use.  Theoretically, the microturbulent velocity should be the same for all 
species.  However in practice, there is sometimes a discrepancy between the 
results measured from different elements and ionization stages.  Of particular 
note, in early type B super giants \ion{Si}{3} sometimes indicates a smaller 
LTE microturbulence value than \ion{O}{2} \citep{mcerlean98,vrancken98}.  \citet{mcerlean99} suggest that in these circumstance the \ion{Si}{3} $4550\mbox{\AA}$ triplet gives a better answer since the lines in the triplet cover a range of strengths and are from the same multiplet, thus less susceptible to systematic errors in the atomic data. 

For this study the microturbulent velocity obtained by balancing the 
\ion{Si}{3} $4550\mbox{\AA}$ triplet has been given preference.  When the 
\ion{Si}{3} $4550\mbox{\AA}$ triplet is not available, a consensus between the other species with the best populated curves of growth is used.  In the cases where fewer than three lines of any species were measured, making it impossible to determine the microturbulent velocity, the value used is ``borrowed'' from another star with a similar effective temperature and gravity.  The microturbulent velocities assigned to each star and the species used to determine them are listed in Table \ref{tab4}.  

The LTE microturbulence values for B stars are typically on the order of 
5 km/s to 10 km/s or more.  Microturbulent velocities of this magnitude 
dominate the broadening parameter for most elements since at 20000 K the thermal 
velocities of hydrogen and iron atoms are 12.8 km/s and 1.7 km/s 
respectively.  Typically the microturbulent velocity can be determined to better than $\pm2$ km/s which translates into an uncertainty in the abundances of less than 0.10 dex.  However, this uncertainty may be higher for stars using ``borrowed'' microturbulence values, especially for the abundances of heavier nuclei with slower thermal velocities where the microturbulence thoroughly dominates the line broadening parameter.  In extreme cases the abundances may be affected by several tenths of a dex.

\subsection{Projected Rotational Velocity}

\begin{figure}
\figurenum{2}
\label{fig1.5}
\includegraphics[angle=90,scale=0.35]{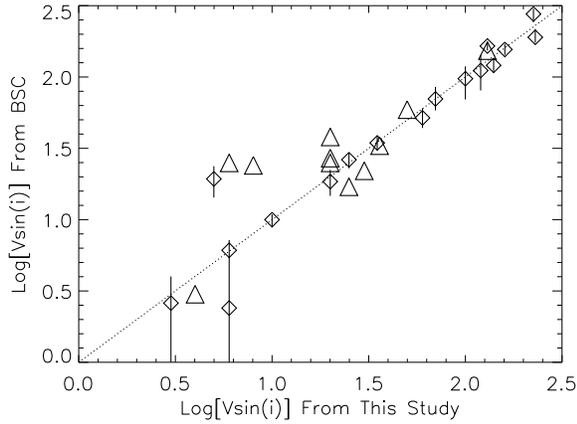}
\caption{A plot of the logarithm of the vsin(i) measured in this study versus the value given in other sources.  The triangles are the nearby B star control sample compared with vsin(i) values from the Yale Bright Star Catalog.  The diamonds are stars from the sample of study compared with values measured by \citet{behr03}.  The vertical lines attached to the diamonds represent Behr's quoted errors.}
\end{figure}

The vsin(i) was determined for each star by synthesizing a piece of spectrum broadened by the method of \citet{gray76} using several different broadening parameters and visually picking the one that best fit the observed profiles.  The \ion{Si}{3} $4550\mbox{\AA}$ triplet or another piece of spectrum containing at least on prominent unblended absorption line were used to determine the best fit vsin(i) for each star listed in Table \ref{tab4}.  The error in each vsin(i) value is about 10\% with a best possible accuracy of about $\pm1$ km/s for the narrowest line stars. 
A comparison of the vsin(i) values measured in this study with other sources shows a good agreement (Figure \ref{fig1.5}).

As part of this work, we wish to distill an sample of Population I runaways which is unbiased with regards to rotational velocity.  Therefore, we are not using vsin(i) as a criterion to classify the stars in the sample.  The distribution of rotational velocities in the sample will be discussed further in Paper II after each star has been assigned a final classification based on both its photospheric abundances and space velocity.

\section{Abundance Analysis}  

\subsection{Stellar Model Atmospheres}

An extensive grid of plane parallel, LTE stellar photospheric models covering 
the range of stellar atmospheric parameters in B stars was calculated using 
ATLAS9 \citep{atlas} at steps of 250 K effective temperature and 0.25 dex 
surface gravity using the standard solar abundance ODFs (Opacity Distribution 
Functions).  We conducted numerical experiments which showed that the metallicity of the ODFs used to calculate the atmospheric models could be changed by up to 2.0 dex and that the constant microturbulent velocity of the models could be varied by up to 15 km/s with no significant effect on the abundance analysis.  The models were calculated using a constant microturbulent velocity of 5 km/s for models hotter than 9000 K and a velocity of 2 km/s for cooler models.  Note that convection can be ignored in most cases since it does not contribute significantly to energy transport in the atmospheric models of B stars.

Uncertainty in the parameters which define the model atmospheres directly 
translate into uncertainty in the abundances.  Each atomic species is affected 
differently by these uncertainties.  Minor species which represent less than 5\% of the population at the level of maximum contribution are very sensitive to 
temperature and pressure.  Therefore, it is minor species which are most affected by perturbations in the structure of the models.  These effects can be quantified by comparing the relative abundances calculated from the same equivalent widths using different models.  
The differences in abundances of selected elements resulting from variation of 
the atmospheric parameters are given in Table \ref{tab5}.  In all cases, the 
minor species show the largest variation with respect to the 
temperature and gravity of the selected model.  On a fairly consistent basis, 
the calculated offsets are well within the typical standard deviation of the 
abundances calculated from the measured equivalent widths.  

\subsection{LTE Radiative Transfer Code}

This study used the LTE radiative transfer codes LINES and MOOG 
\citep{sneden73, luck01b}.  LINES is used to convert equivalent widths for 
single transitions into abundances while MOOG is used to synthesize and fit 
line profiles with one or more components.  Both codes were optimized to perform abundance analysis in the B star temperature regime.  The dominant line broadening mechanism in B stars is the quadratic Stark effect.  Our experiments showed that the \citet{cowley71} method is best able to reproduce the observed line profiles so it was installed as the method used to calculate the Stark broadening parameters in both codes.

Both MOOG and LINES passed a number of consistency checks.  
A piece of spectrum synthesized by MOOG is comparable to 
the same spectrum synthesized by SYNTHE \citep{atlas}.
LINES is able to successfully recover the abundances for equivalent widths synthesized by MOOG as well as reasonably reproduce the abundance analysis results of \citet{gl92} and \citet{hambly97} from the data given in those publications.

\subsection{LTE versus NLTE}

NLTE effects, which are normally characterized as atomic level population 
perturbations relative to LTE, influence both individual line formation and 
the overall structure of a stellar atmosphere.  These effects are more 
significant for some lines than others.  \citet{kilian94} found that at the $\pm$(0.1 to 0.2) dex level there is no difference between the absolute 
abundances calculated using LTE assumptions versus NLTE calculations.  
Whereas, \citet{daflon01a} found that the difference between LTE and NLTE abundances in B stars depend on the atomic species.  They showed that differences between LTE and NLTE derived abundances of \ion{C}{3} and \ion{Al}{3} are negligible while \ion{N}{2}, \ion{O}{2}, and \ion{Si}{3} NLTE abundances are 0.1 to 0.2 dex higher than LTE abundances and \ion{Mg}{2} NLTE abundances are up to 0.25 dex lower than LTE abundances.  However, these differences are less significant in a relative abundance analysis where the abundances of the control sample are affected in the same manner as the sample of study.  

\subsection{Line and Atomic Data}

The atomic transition data used in this analysis was obtained from the 
NIST online atomic 
database\footnote{http://physics.nist.gov/PhysRefData/contents-atomic.html} 
and the sources listed therein.  The lines used in the analysis 
were selected from the measured absorption features as the ones that do not
show large a abundance spread in the nearby Population I control sample (see the next section).  The lines and atomic data used to model each transition are listed in Table \ref{tab8} (available only in the electronic version of this paper).

\subsection{Abundances of the Population I Control Sample}

The absolute abundances measured for the stars in the Population I control sample are given in Table \ref{tab6}.  The abundances of each element are measured using the dominant species in each star.  The average abundances of this control sample compare favorably with the results from a variety of other studies (also Table \ref{tab6}).  Note that the observed abundances of some elements in B stars are depressed relative to the established solar values.  We assure the reader that this is a universally observed phenomenon which should not cause undue concern especially in the case of a relative abundance analysis (see references at the bottom of Table \ref{tab6}).
 
\subsection{Abundances Measured Relative to the Control Sample}

It is easiest to measure the relative abundances between two stars with the same atmospheric parameters because then the only variable which they do not have in common is their composition.  In a perfect universe there would be one star that we could observe as a standard which all other stars could be compared to (i.e. the Sun).  However in reality, stars cover such a large range of effective temperatures that the atomic transitions and processes in a star on the cool end of the scale have almost no application in a star on the hot end.  Unfortunately there is even a large range of physical conditions in B stars ($T_{eff}$= 10000 K to 35000 K).  The spectrum of an early type B star bears almost no resemblance to a late type B star, save for the ever present hydrogen and helium lines.  In order to do a good job of measuring the relative compositions of B stars, one must find a way to tie the abundances of these stars together over the wide range of parameter space that they occupy.  In this study, this is accomplished by building an ``average star'' from the control sample.  Relative stellar abundances are calculated on a line by line basis as the difference with respect the average absolute abundance measured for that line in the Population I control sample (Table \ref{tab8} with the relative program abundances in Table \ref{tab9}). 

The control sample of nearby Population I B stars is drawn from a population that is for the most part chemically homogeneous \citep{kilian92,gl92,daflon03}.  
In the case of a homogeneous population we can assume that differences in the equivalent width measured for a given line in each of these stars are not due to differences in composition but primarily caused by differences in the structure of the stellar atmosphere as defined by the photospheric parameters.  Therefore, transitions that give abundances which show trends with respect to temperature or large random spreads are dropped from the analysis as unreliable indicators of relative abundance.  

In the course of generating a list of reliable atomic transitions from the nearby Population I control sample in this manner, it became apparent that some atomic species were reliable indicators of abundance over part of the temperature range, but then became less reliable outside of that range.  To compensate for this, the rules in Table \ref{tab7} were adopted to restrict the use of lines from some species only over a specified range of effective temperatures.

Some lines were measured in only one star in the nearby Population I control sample.  The lines in this situation which yield abundances that are consistent with other reliable lines in that star were given the benefit of the doubt and included in the line list.  In a perfect world there would have been enough lines to choose from that one would not have to take a chance that some of these lines measured in only one star were unreliable.  However there are not a large number of spectral lines in most B stars so it is pragmatic to keep a line in the list until it has proven itself unreliable.    

Ideally this method should reduce the standard deviation about the mean abundance measured from several individual lines in the same star.  Unfortunately, in practice the promise of the method is not always achieved.  The difference between the standard deviations from the absolute analysis subtracted from the relative analysis are given in Table \ref{tab10}.  There is a significant improvement in average standard deviations of six species (\ion{Al}{3}, \ion{Si}{3}, \ion{Ti}{2}, \ion{Cr}{2}, \ion{Fe}{2}, \ion{Fe}{3}).  Those species most improved either are measured from a small list of very well behaved transitions or are only used over a narrow temperature range.  In either of those cases the conditions are most ideal for relative abundance analysis to remove effects from the atomic data and atmospheric parameters to first order.  All but one of the other species showed no significant change in precision as a result of relative abundance analysis.  The average standard deviation of  \ion{Si}{2} in the relative abundance analysis is significantly larger than the value computed in the absolute abundance analysis because \ion{Si}{2} is very sensitive to changes in temperature and pressure structure of the atmosphere.  Therefore \ion{Si}{2} is not well suited for relative analysis and when possible \ion{Si}{3} was used instead to determine silicon abundance. 

\begin{figure}
\figurenum{3}
\label{fig2}
\plotone{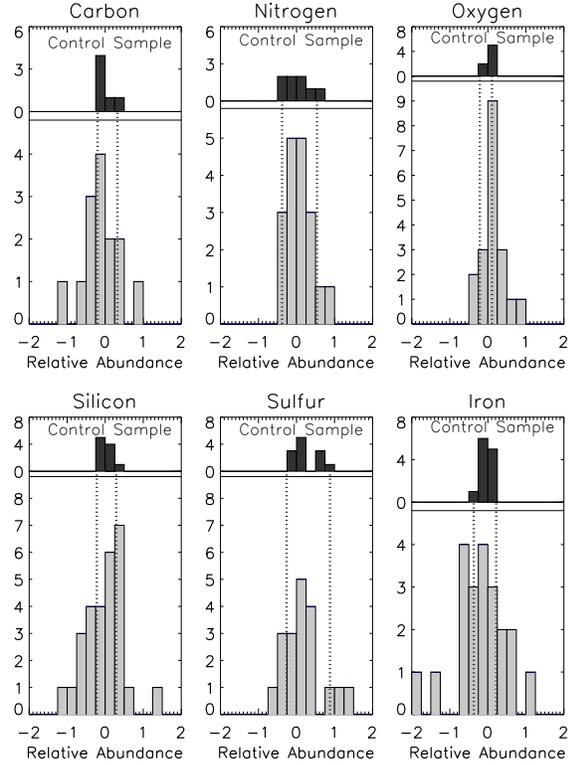}
\caption{These are histograms of the relative abundances of carbon, nitrogen, oxygen, silicon, sulfur, and iron for the sample of high latitude B stars.  The nearby Population I control sample is plotted at the top of each histogram for comparison.  The dotted vertical lines are the minimum and maximum value in the nearby Population I control sample.}
\end{figure}

The elemental ratios and discussion of elemental abundances are based on the major ionization species, which is often the one with the most lines measured.  Figure \ref{fig2} shows the distribution of elemental abundances in the sample of study measured relative to the control sample.

\subsection{Abundance Ratios}

\subsubsection{Selecting Referents}

Abundance ratios are helpful for determining the pattern of enhancements and depletions in a stellar atmosphere.  Of primary interest to this study are enhancements of carbon, nitrogen, oxygen, s-process, $\alpha$-process, and refractory elements which are used to distinguish between BHB, PAGB, PPN, and main sequence stars.  In order for a ratio to effectively detect an enhancement, depletion, or lowered measured abundance of an element it must be compared to another element that is unaffected by the process for which one wishes to test.

An element which is unaffected by all of these processes would be the ideal universal referent.  Pragmatism dictates that the referent selected must be measurable over the same range of atmospheric parameters as the element that is being tested for variation.  However, it has proven difficult to identify one element with all these qualities so instead several different elements are used as referents, each with its own strengths and weaknesses.  

Iron is often used as a referent.  However, it can be affected by dust grain formation (especially in PAGB stars).  Also, iron has few prominent lines in hotter stars making it difficult to measure its abundance, especially in stars with significant vsin(i).   Table \ref{tab5} demonstrates that minor species of iron tend to be more affected by small changes in the temperature and pressure structure of the stellar atmosphere.  Additionally, iron's high atomic weight gives it a smaller thermal velocity, which makes it more sensitive to changes in microturbulence.   

Aluminum is a good referent because it remains untouched by most of the nuclear processes that affect the photospheric composition of stars through the PAGB and PPN phase.  Also, Al III lines are prominent and easy to measure in hot B stars.  However, aluminum is affected by the initial composition of the star \citep{edvardsson93,tomkin85} and its measured abundance can be lowered by dust formation.  The effect of initial composition is slight and will not hinder the ability to measure significant enhancements in other elements due to most nuclear processes.  The influence of dust formation on the aluminum abundance can be more significant.   However, aluminum is still a good referent as long as one keeps in mind that its abundance may be lowered in some stars which can be identified by other means. 

Oxygen is a good referent to use with carbon and nitrogen to test for the presence of CNO processed material.  It is affected less by that process than carbon or nitrogen and the oxygen abundance may be measured from many prominent O II lines over the entire range of stellar atmospheric parameters in B stars.  However, one should be aware though that there are processes in later stages of stellar evolution that might enhance oxygen abundances \citep{imbriani01}.

Sulfur is the best element in this study to measure the lowering of abundances due to grain formation.  However, it is an $\alpha$-process element which is also affected by initial composition.  This effect may cancel out to first order if sulfur is used as a referent to measure the lowering of the abundance of another $\alpha$-process element, like silicon.

\subsubsection{Ratios in the Program Stars}

\begin{figure}
\figurenum{4}
\label{fig3}
\plotone{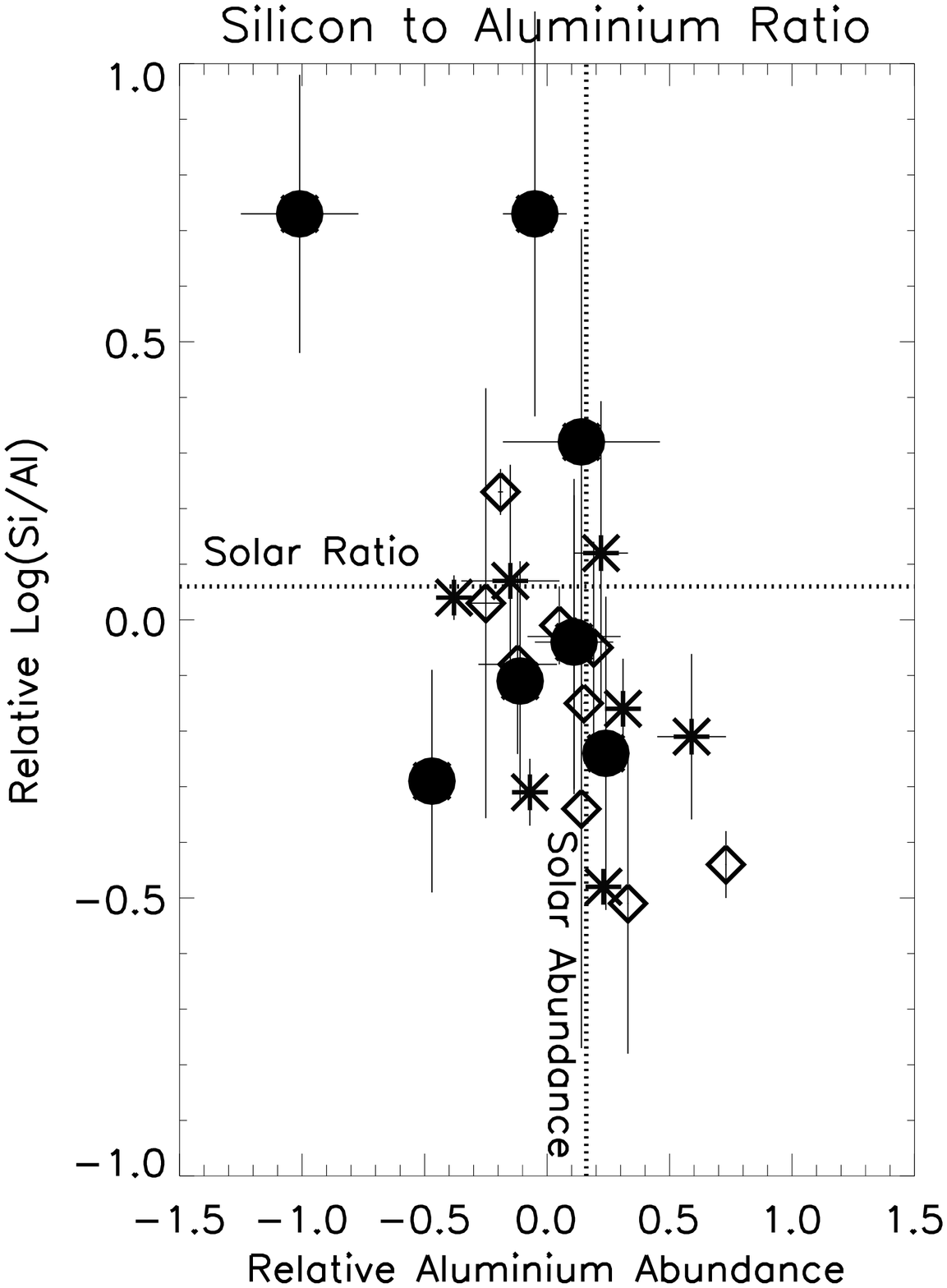}
\caption{The diamonds are the nearby Population I control sample.  The solid circles and *'s are the program stars which respectively have and do not have more than one abundance ratio which are more than two standard deviations from the mean of the control sample (see section 4.7.2).}
\end{figure}

\begin{figure}
\figurenum{5}
\label{fig4}
\plotone{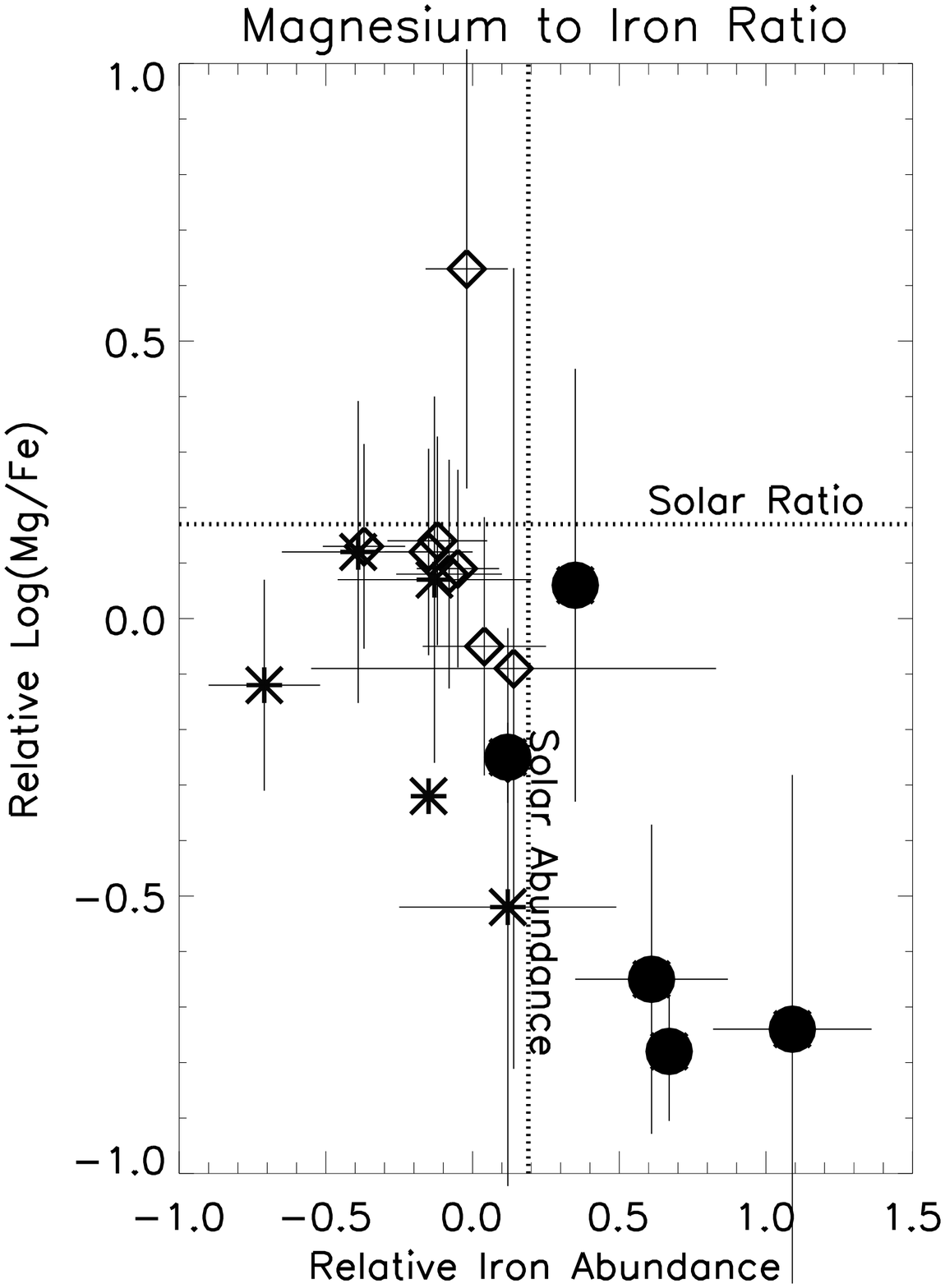}
\caption{The diamonds are the nearby Population I control sample.  The solid circles and *'s are the program stars which respectively have and do not have more than one abundance ratio which are more than two standard deviations from the mean of the control sample (see section 4.7.2).}
\end{figure}

\begin{figure}
\figurenum{6}
\label{fig5}
\plotone{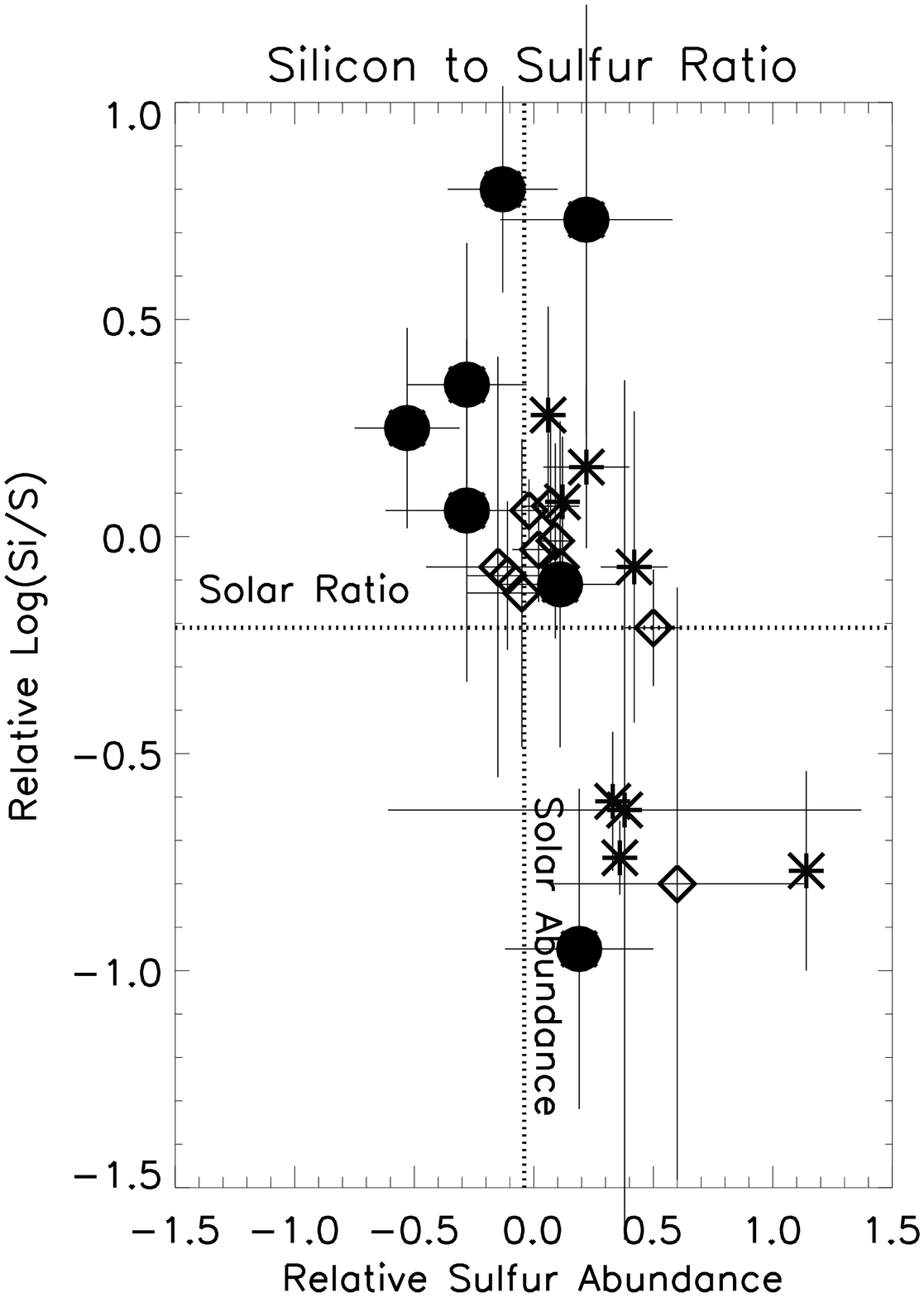}
\caption{The diamonds are the nearby Population I control sample.  The solid circles and *'s are the program stars which respectively have and do not have more than one abundance ratio which are more than two standard deviations from the mean of the control sample (see section 4.7.2).}
\end{figure}

\begin{figure}
\figurenum{7}
\label{fig6}
\plotone{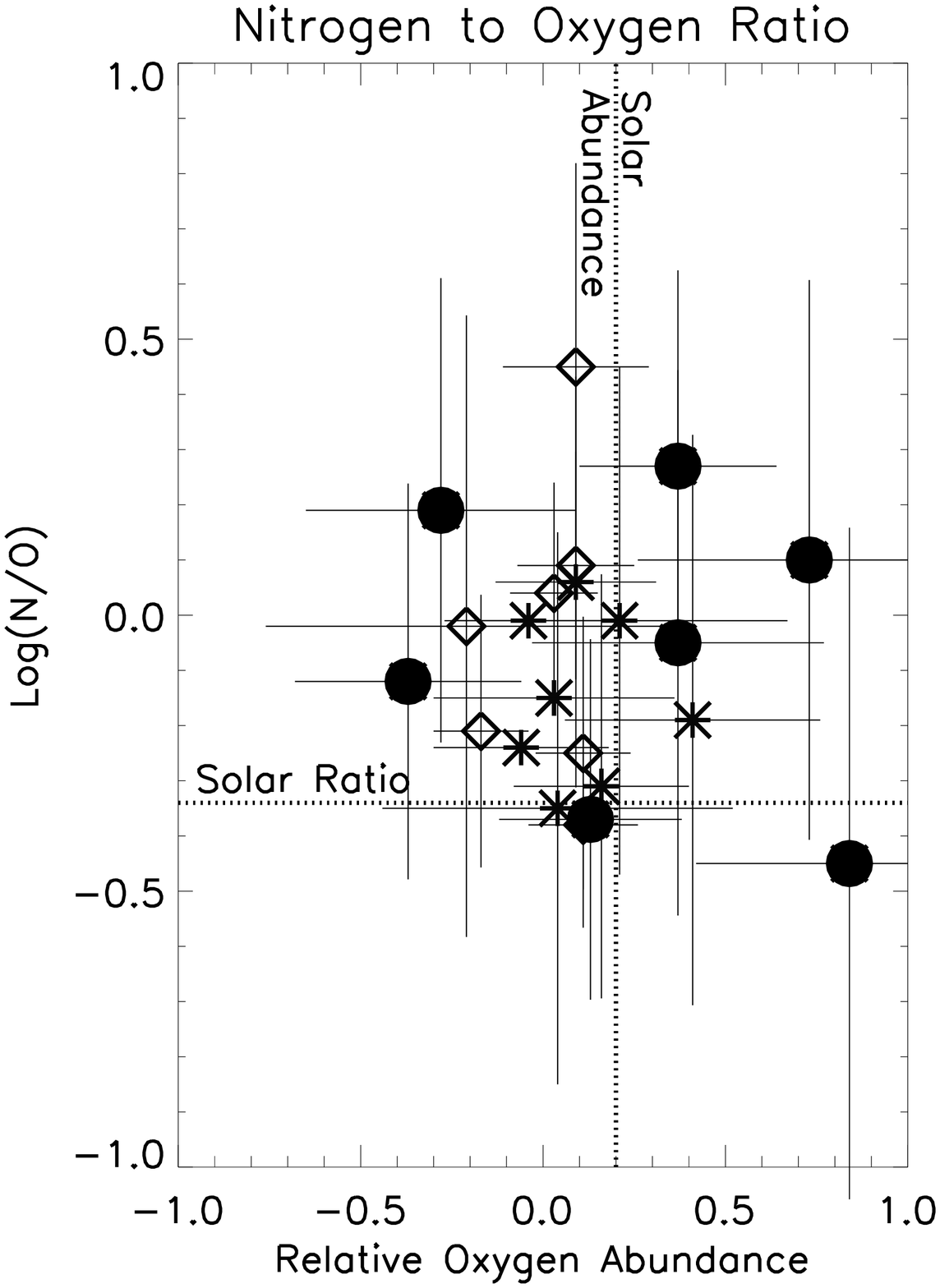}
\caption{The diamonds are the nearby Population I control sample.  The solid circles and *'s are the program stars which respectively have and do not have more than one abundance ratio which are more than two standard deviations from the mean of the control sample (see section 4.7.2).}
\end{figure}

\begin{figure}
\figurenum{8}
\label{fig7}
\plotone{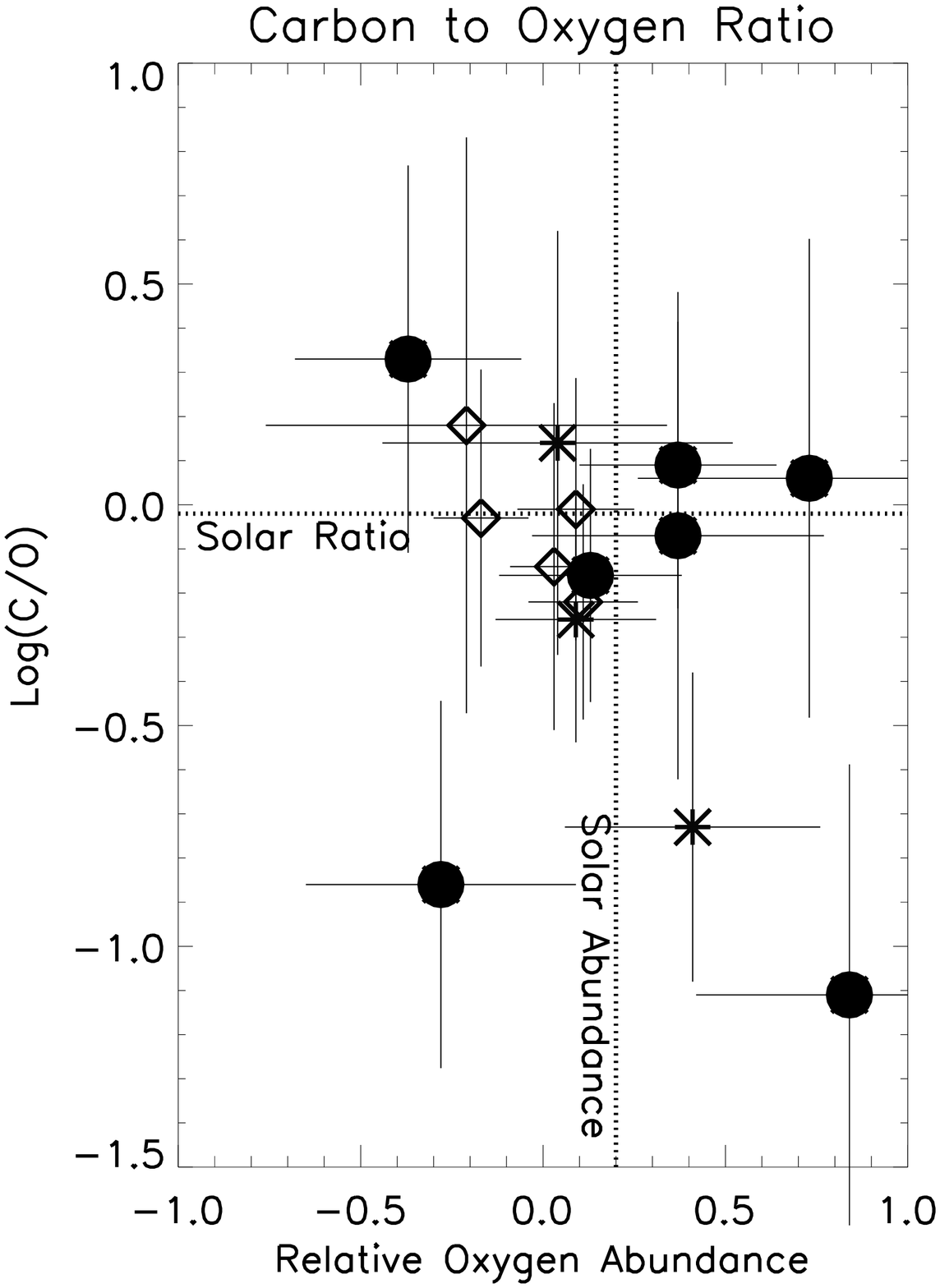}
\caption{The diamonds are the nearby Population I control sample.  The solid circles and *'s are the program stars which respectively have and do not have more than one abundance ratio which are more than two standard deviations from the mean of the control sample (see section 4.7.2).}
\end{figure}

\begin{figure}
\figurenum{9}
\label{fig8}
\plotone{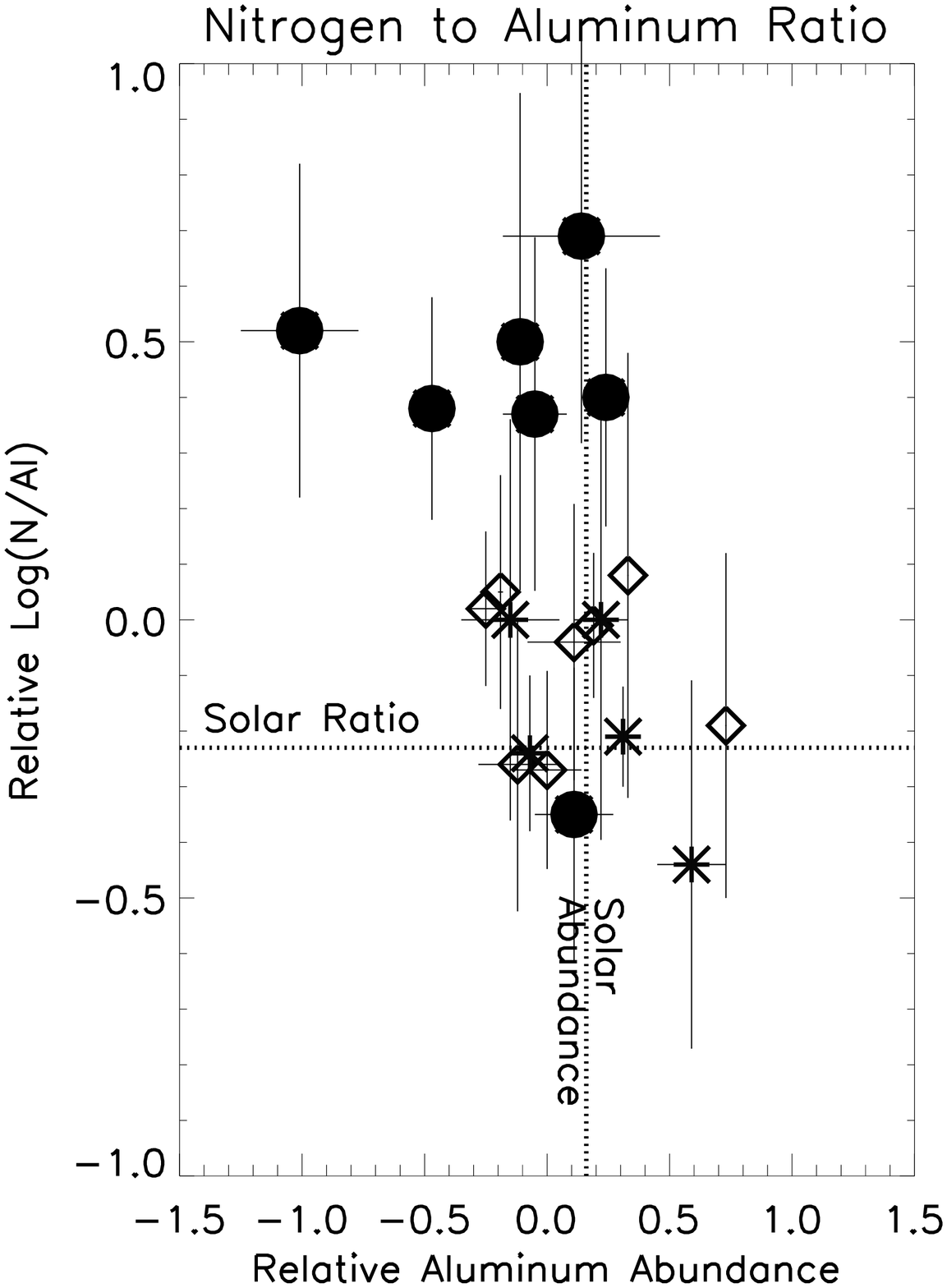}
\caption{The diamonds are the nearby Population I control sample.  The solid circles and *'s are the program stars which respectively have and do not have more than one abundance ratio which are more than two standard deviations from the mean of the control sample (see section 4.7.2).}
\end{figure}

\begin{figure}
\figurenum{10}
\label{fig9}
\plotone{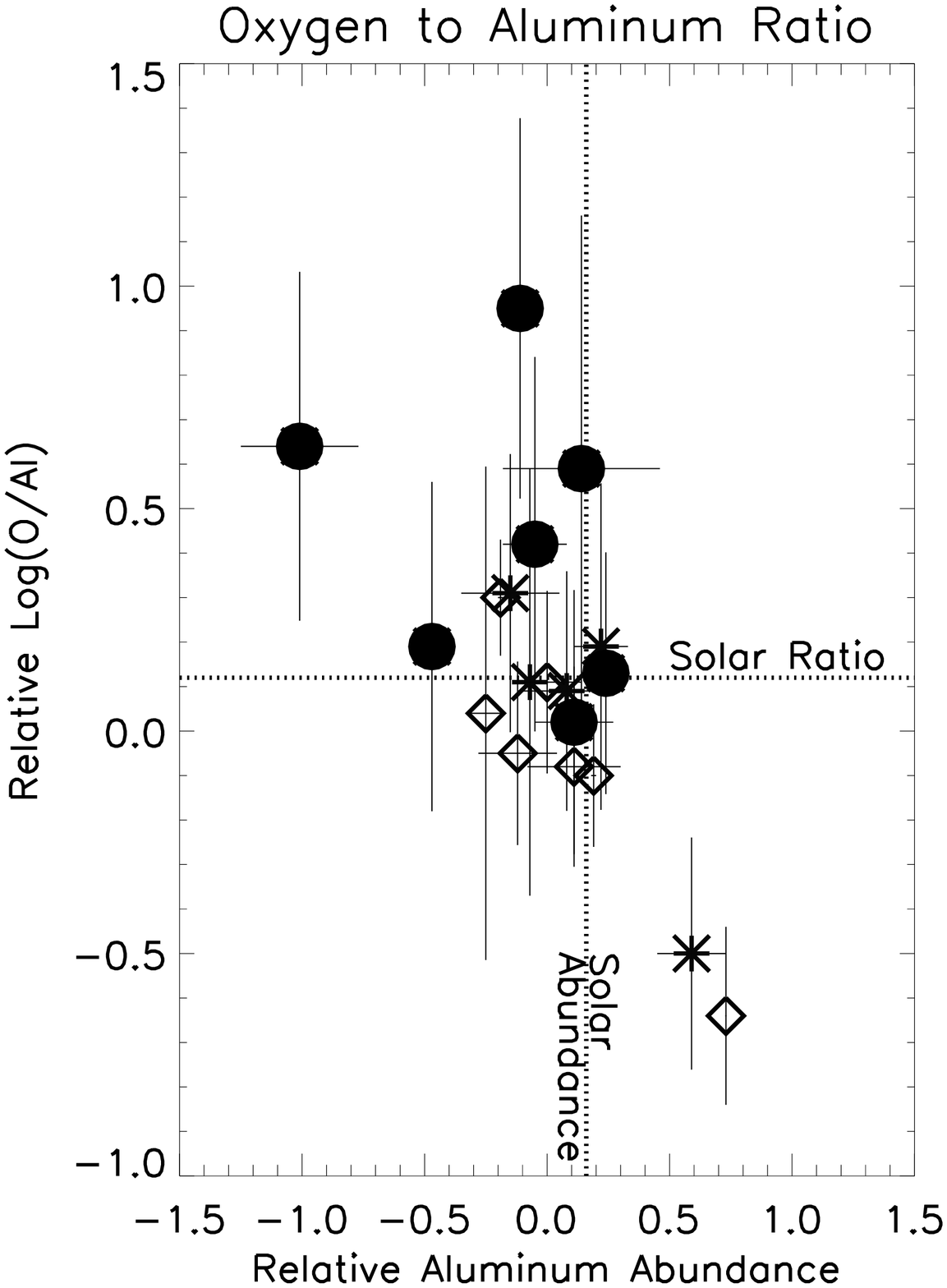}
\caption{The diamonds are the nearby Population I control sample.  The solid circles and *'s are the program stars which respectively have and do not have more than one abundance ratio which are more than two standard deviations from the mean of the control sample (see section 4.7.2).}
\end{figure}

The abundance ratios measured in the program stars and BHB and PAGB control samples are in Table \ref{tab9}.  Three ratios, Si/Al (Figure \ref{fig3}), Mg/Fe (Figure \ref{fig4}), and Si/S (Figure \ref{fig5}), are used in this study to detect the influence of initial composition and dust formation on the measured abundances.  Meanwhile, the N/O (Figure \ref{fig6}), C/O (Figure \ref{fig7}), N/Al (Figure \ref{fig8}), and O/Al (Figure \ref{fig9}) ratios are used to identify enhancements that could be due to mixing of processed material into a stellar atmosphere.  These ratios must be interpreted keeping in mind that they involve species which are affected by the temperature and pressure structure of the atmosphere.  Therefore, in some cases these ratios can be more sensitive to the assumptions made about the atmosphere than the composition.

For each abundance ratio the distribution of values in the nearby star control sample (plotted as open diamonds in Figures \ref{fig3} through \ref{fig9}) is used to determine the usual distribution of values found among Population I B stars in the disk.  Exceptional abundance ratios for program stars (more than two standard deviations from the mean of the control sample) are italicized and underlined in Table \ref{tab9}.  Seven program stars (BD -15 115, HD 21305, HD 40276, HIP 41979, BD +36 2268, PB 166, and BD +33 2642) have more than one exceptional abundance ratio.  HD 21305, HIP 41979, and PB 166 are probably BHB stars (see section 4.9) and BD +33 2642 has been classified as a proto-planetary nebula with abundances which have been affected by an initial metal poor halo composition and dust formation \citep{napiwotzki94,napiwotzki01}(see Appendix A).  These stars which have abundance ratios that are persistently different from the nearby Population I control sample (plotted with solid circles in Figures \ref{fig3} through \ref{fig9}) are the best candidates to be old low mass evolved stars.  Eight more program stars (HD 21532, BD +61 996, HD 233622, HD 105183, HD 146813, HD 149363, HD 209684, and HD 218970) have one exceptional ratio each.  The argument could be made to also classify these eight stars as low mass evolved stars based on their abundance ratios.  However, their status is less certain.

\subsection{Classification Based on Abundances}

The results of the abundance analysis divide the sample of high galactic latitude B stars into three categories which are given in Table \ref{tab9}:  disk, metal poor (MP), or old evolved star (OES).  The results for the BHB and PAGB star control samples are also presented in Table \ref{tab9} for comparison.  Unfortunately it was not possible to obtain abundances for every element in each star so in some cases the picture presented by the relative abundances and ratios is incomplete.  In those cases the uncertainty has been noted through a question mark (?) after the classification.  The ``Disk'' category contains twenty two (22) stars which have abundances and ratios consistent with nearby Population I B stars.  Presumably these stars are the best candidates for consideration as runaway stars ejected from the disk.  The ``old evolved star'' category includes nine stars which have a pattern of enhancements and depletions of key elements which could be the result of advanced stellar evolution.  Eight stars can only be characterized as ``metal poor'' (MP) with abundances which are uniformly lower than the control sample.  Those stars could potentially turn out to be old evolved stars or stars which formed in situ in the galactic halo.

\subsection{Metal Rich BHB Stars}

Two stars in the sample (HIP 41979 and  PB 166) have atmospheric parameters, abundance patterns, and projected rotational velocities which are consistent with hot horizontal branch stars.   Each of these stars have CNO ratios consistent with post main sequence evolution and mixing as well as high iron abundances which are be indicative of a purely radiative BHB atmosphere affected by diffusion \citep{michaud83}.  Curiously, the magnesium abundance, which is unaffected by radiative diffusion \citep{behr99,behr00}, is solar in these stars.  This is surprising since metal rich populations should not produce BHB stars \citep{iben70,rood73} and surveys of BHB stars in the solar neighborhood have not found any with abundances above 1/10 solar \citep{kinman94,kinman00}.  Therefore, these BHB stars are a significant find, especially since they are much nearer and brighter than the others discovered to date \citep{liebert94,peterson01,peterson02}.

HD 21305 has an abundance pattern and projected rotational velocity which are also consistent with a solar metallicity BHB star affected by diffusion.  However, the atmospheric parameters of HD 21305 ($T_{eff}$ = 18500 K; log(g) = 4.0) are more consistent with a main sequence or PAGB star since its surface gravity is about 0.4 dex above the terminal age horizontal branch (TAHB).  The analysis of this star's kinematics in Paper II will be the decisive factor which will determine if it is the third metal rich BHB in the sample.

\section{Other Data from Spectra}

\subsection{Strength of Helium Lines}
The abundance of helium can be affected by stellar evolution.  For example, some older evolved stars have helium abundances which are higher than those found in ``normal'' Population I B stars.  On the other hand, radiative diffusion causes the the equivalent widths of \ion{He}{1} lines in BHB stars to appear weaker than main sequence stars with solar abundances \citep{greenstein67}.  

\begin{figure}
\figurenum{11}
\label{fig12}
\includegraphics[angle=90,scale=0.35]{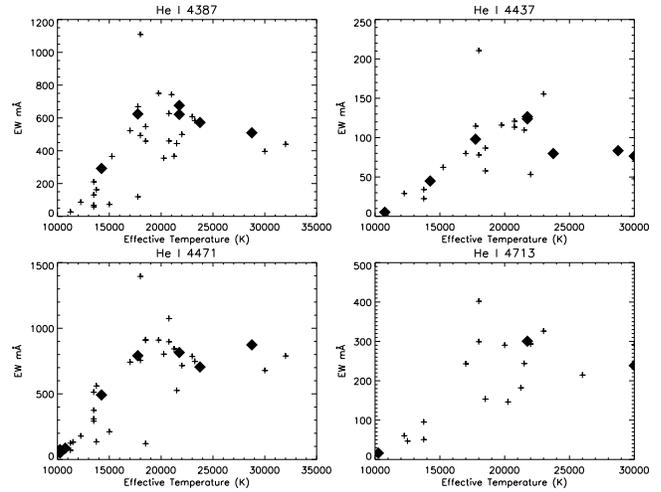}
\caption{The equivalent widths of four He I lines versus effective temperature.  The equivalent widths of the nearby Population I control sample are the black diamonds.  The +'s are the program stars.}
\end{figure}

The neutral helium lines in the spectra of B stars can be heavily influenced by NLTE factors and multiple blended components which make it difficult to calculate absolute abundances.  Relative helium abundances can be determined by comparing the equivalent widths of lines in 
two stars with the same atmospheric parameters.  Figure \ref{fig12} shows that almost all the high latitude B stars and the nearby Population I B stars fall on the same distribution as a function of effective temperature, indicating that they have similar helium abundances.  There are a few stars with equivalent widths that are significantly larger or smaller than the control sample.  

HIP 41979 (the high point at 18000 K) has consistently larger equivalent widths than the rest, indicating that it could possibly be helium enriched.  BD +13 3224 (V625 Her) is also overabundant in helium \citep{jeffery01}.  However, it has not been plotted in Figure \ref{fig12} because its short pulsational period and rapidly changing photosphere conspired with the long exposure times in this study to render the spectrum recorded for it unmeasurable.  Meanwhile, HD 21305, HD 105183, and PB 166 have equivalent widths which are consistently smaller than the control sample, indicating that they could possibly be BHB stars affected by diffusion.  

In addition to having abnormal \ion{He}{1} equivalent widths, HD 21305, HIP 41979, and PB 166 also have abundance patterns which suggest that they are metal rich BHB stars (see Section 4.9).  It is unusual that HIP 41979 has larger than normal \ion{He}{1} equivalent widths since BHB stars affected by diffusion should appear to be deficient in helium (leading to smaller than normal equivalent widths).  \citet{ramspeck01a} identified three BHB stars with a similar abundance patterns which also appear to have elevated helium abundances.  It appears that HIP 41979 also fits into this group of BHB stars which without clear explanation defy conventional wisdom concerning the mechanics of radiative diffusion in BHB atmospheres.

\subsection{Emission Features}
Exposures were taken in sets so that emission features, which appear in 
every exposure, could be distinguished from cosmic ray hits.  Great care was taken during processing to identify any emission features in the observed spectra.  Spectral emission features were only found in two of the sample stars (BD +33 2642 and Feige 23).  

BD +33 2642 is a previously identified halo proto-planetary nebula 
(PPN) \citep{napiwotzki93,napiwotzki94}.  It has an unusual spectrum composed of a blend between classic nebular emission lines and a stellar photospheric
absorption spectrum, which allows a unique opportunity to analyze the 
abundances of both the nebula and the central star.  This analysis and a more extensive discussion of this object are in presented in Appendix A.

\begin{figure}
\figurenum{12} 
\label{fig13}
\includegraphics[angle=90,scale=0.35]{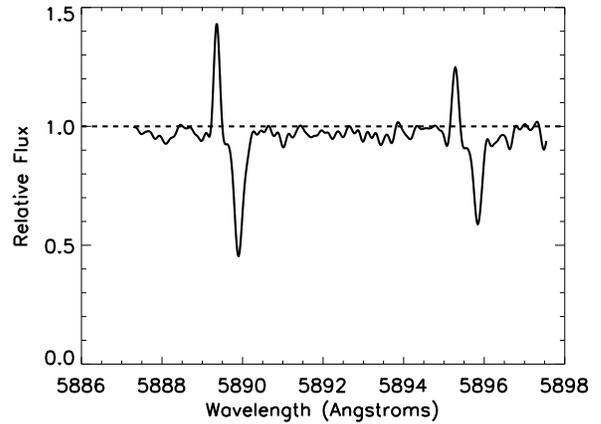}
\caption{The sodium D lines in emission in the spectrum of Feige 23.}
\end{figure}

Feige 23 has a normal spectrum with very narrow sodium D emission lines (Figure \ref{fig13}).  This emission, combined with inconsistencies in the (B-V) colors measured for
this star \citep{klemola62,blanco70,kilkenny95,hip} are probably indicative of some kind of dust shell or other aggregation of circumstellar material.

\subsubsection{Be Stars}
Curiously, there are no stars in this sample which exhibit the classic emission features used to identify Be stars.  The frequency of Be stars among nearby stars \citep{zorec97} leads to the expectation that a sample of this size and distribution of spectral types should have about ten Be stars.  It is important to note that the number of Be stars in a sample will almost always be underestimated because Be stars do not always exhibit emission features.  Also, only the runaway Population I B stars in the sample are potentially drawn from the same population as the nearby stars surveyed by \citet{zorec97}.  Even taking these factors into account it seems unlikely that the sample would have no identifiable Be stars in it unless somehow the ejection mechanisms for runaway B stars select against stars exhibiting the Be phenomenon.  

Be stars tend to rotate faster than the general population \citep{slettebak82} and are also thought to be created through interaction with a close companion in a binary system (for a review and references see \citet{porter03a}).  Therefore, a population with few fast rotators and no binary companions might have few if any Be stars.  Runaway B stars are far less likely to have binary companions than B stars in the disk field population \citep{gies86}.  However, it will be shown in Paper II that the average vsin(i) of runaway B stars is 
{\em faster\em} than the average vsin(i) of nearby disk B stars.  Is it possible that the lack of binary companions by itself could be responsible for the absence of Be stars from our sample?  We do not know.  However, further study of high latitude B stars may yield more information about the origin of the Be phenomenon.

\section{Conclusions}

\subsection{Preliminary Classification of the Sample}

Detailed photospheric elemental abundance analysis of forty program stars has revealed that while twenty two stars in the sample have abundances which appear to be indistinguishable from nearby Population I B stars, nine others have abundance patterns which strongly indicate that they are old evolved stars, eight have abundance patterns which are metal poor, and one (HD 1112) has an abundance pattern which we are unable to classify.  We also found that AG +03 2773 is probably an rapidly rotating F dwarf which has been misclassified as a B star.  The lines in the spectrum of AG +03 2773 are broad and numerous so that they effectively obliterate the continuum at most wavelengths and make it difficult to perform an abundance analysis.

In addition to the abundance data, we found that seven stars in our sample have spectral peculiarities which are best interpreted by phenomenon which are not expected in runaway Population I B stars:  \ion{He}{1} equivalent widths outside the range in ''normal'' B stars of a 
given temperature, and/or odd spectral emission features.   Five of these stars also have abundance patterns presented here which verify that they are probably old evolved stars and one of the remaining two (BD +13 3224) is a previously identified pulsating helium dwarf.

Table \ref{tab12} gives a preliminary classification of each star.  The sample is broken into four categories: 
\begin{itemize}
\item{Population I runaway (Pop I; 21 stars),}
\item{Old evolved BHB or PAGB star (OES; 10 stars),} 
\item{F dwarf (1 star),} 
\item{or unknown (17 stars).}
\end{itemize}
The category ``unknown'' covers stars with insufficient or conflicting information which preclude their placement in one of the other categories.  Therefore at this stage, ``unknown'' includes any B stars formed in situ in the halo since we require an analysis of a star's kinematics and past trajectory to make that classification with certainty.  These classifications were made conservatively in order to minimize the number of stars that might migrate from the Population I to the old evolved star category when the results from the kinematic data are considered in Paper II.

\subsection{Comparison with Other Work}
A number of the stars in our sample have been categorized individually or in small groups by other studies.  In most cases those classifications agree with the classifications made here.    This includes:  BD -15 115 \citep{ramspeck01b,magee01}, BD +38 2182 \citep{conlon89}, HD 100340 \citep{ryans99}, HD 105183 \citep{dufton93}, Feige 40, PB 166 \citep{deboer88}, Feige 84 \citep{lynn04,saffer97}, HD 123884 \citep{bidelman88}, HD 137569 \citep{danziger70}, HD 138503 \citep{itlib}, BD +33 2642 (See Appendix I), and BD +13 3224 \citep{jeffery01,jeffery86}.  

Two of the stars in this study have received classifications which potentially conflict with the results from other assorted studies.  \citet{conlon92} classified HD 121968 as a normal Population I B star.  We have no spectrum to measure the abundances for this star.  Indications are that it is probably a runaway.  However, we are reserving judgment until after we have analyzed its kinematics so for the moment it remains classified ``unknown.''  

\citet{zboril00} classified HD 149363 as a slightly helium rich evolved star with carbon enhancement but normal nitrogen and oxygen abundances.  However, we classify HD 149363 as a Population I runaway because unlike \citet{zboril00} we observe no abnormality in the strength of the helium lines and measure carbon, nitrogen, and oxygen abundances which do not differ significantly from stars in the nearby Population I control sample.  \citet{zboril00} used nearly the same temperature and gravity to analyze this star.  Therefore, the difference between our results is probably due to the microturbulence and/or the atomic transitions measured.  \citet{zboril00} adopted a microturbulence value of 6km/s while this work measured a microturbulent velocity of 20 km/s.  If the microturbulence is reduced to 6 km/s in our analysis then the abundances of carbon, nitrogen, silicon and sulfur rise by 0.3 dex, aluminum abundance rises by 0.4 dex, and oxygen abundance rises by 0.5 dex.  This does not completely match the pattern of abundances which they reported.  In addition to the difference in microturbulence, \citet{zboril00} use of the C II 4267 \mbox{\AA} doublet alone to measure carbon abundance.  It is well established that this feature is an unreliable measure of absolute carbon abundance due to very strong NLTE effects \citep{sigut96,kane80}.  \citet{zboril00} argue that these effects should lead to a systematic underestimate of carbon abundance rather than an over abundance.  However, with the difference in microturbulence included, the behavior of this transition may account for the difference between our analyses.

Fifteen stars in our sample were also cited by \citet{conlon92} as stars with full space velocities which imply that they were ejected from the galactic disk.  We have classified all these stars, with the exception of BD +36 2268, as either Population I runaways (nine stars) or stars which we are reserving judgment about until we have analyzed their space velocities (five stars).  BD +36 2268 may well have a space velocity consistent with ejection from the galactic disk but it also has several abnormal abundance ratios (see Table \ref{tab9}) which imply that is is an old evolved star.  In this case, the abundance analysis should be given precedence because kinematics alone are unable to determine which stars are old evolved post-main sequence objects.

\citet{behr03} classified 21 stars in this study according to their position relative to evolutionary tracks on an $T_{eff}$ versus log(g) plot.  The effective temperatures and surface gravities derived by Behr for these stars closely match the values which we have adopted in this study.
\citet{behr03} also measured the LTE abundances of iron and magnesium relative the solar values.  All but one of the iron and magnesium abundances measured in this study are consistent with Behr's values within the quoted errors.  (The discrepancy in the iron abundance of BD +30 2355 is discussed below.)  There are also no conflicts between our classifications and Behr's for 16 of the 21 common stars.  The difference in classification for the remaining five stars mostly arises from the consideration of additional information which \citet{behr03} did not incorporate into their study.  

HD 1112, HD 233622, HD 105183, and BD +36 2268 (classified "main sequence" by \citet{behr03}) are classified as OES or ``unknown'' here because they have abundance patterns which are not consistent with the nearby Population I star control sample.  This study analyzes the abundances of several light elements, including carbon, nitrogen, and oxygen, which are affected by stellar evolution while \citet{behr03} only considered the abundances of iron and magnesium in the context of matching stars to evolutionary tracks.  In addition to having a pattern of enhancements and depletions of specific elements which is indicative of post-main sequence evolution, HD 105183 also has weaker than expected helium lines, which is more typical of an OES.  

BD +30 2355 was classified by \citet{behr03} as a ``possible'' BHB star.  Whereas, we classify it as Pop I.  In the discussion of their results \cite{behr03} expressed some reservations about their own classification for this star since its vsin(i) is rather large (100 km/s) for a BHB star.  While Behr measured the [Fe/H] of BD +30 2355 to be $-2.70\pm35$ from an unspecified number of lines we found that its iron abundance was only $-0.22\pm.38$ dex relative to the Population I control sample (measured from two lines).  There is a difference in the microturbulence values which each of us adopt (2km/s for Behr and 5km/s for us).  However, lowering the microturbulence to 2 km/s in our analysis {\it raises} the iron abundance by almost one dex.  In final analysis, the source of the disagreement between this study and \citet{behr03} is unclear.  Since we have measured a disk-like metallicity and no other spectral peculiarities for BD +30 2355, we have given it a preliminary classification of Pop I, pending verification by examination of its space velocity in Paper II.

\subsection{Other Results}

Among the stars in the sample with old evolved star abundances, there are as many as three solar metallicity BHB stars.  These metal rich BHB stars are much closer and brighter than the others discovered to date.  We are reasonably certain that HIP 41979 and PB 166 are correctly identified as metal rich BHB stars.  However, conflicting information about HD 21305 means that its classification depends on the kinematic data to be presented in Paper II.

It is surprising to find that there are no active classical Be stars in this sample.  While there is a slight chance that small number statistics may be responsible, this seems unlikely.  The only difference we can find between runaway B stars and Be stars (other than the defining emission features) is the low incidence of binary companions among runaways.  Further study is required to tell whether or not this is the factor responsible for the missing Be stars in our sample.  

In Paper II we will present the kinematic data for the stars in the sample.  We will use this data and simulations to show that the space velocities of runaways retain a distinct kinematic signature which can be used to separate them from halo stars.  We will also calculate and compare the flight times and main sequence lifetimes of the stars in the sample in order to determine which of them could have originated in the disk.  That data will be combined with the data from this paper to classify each star as a Population I runaway, an old evolved star, or a star formed in situ in the halo.  With a sample of Population I runaways in hand we will analyze their distribution of ejection velocities and projected rotational velocities in order to gain insight into the mechanisms which ejected them from the disk.  Additionally, we will also analyze the vsin(i) of PAGB stars and the kinematics and origins of the BHB stars identified in the sample.  

\section{Acknowledgments}
I want to thank my mentor and friend R. Earle Luck for his tireless support of my research and his continued thoughtful questioning and input.  Apparently an advisor's work is never done.  I also want to thank the rest of my dissertation committee, Heather Morrison, J. Chris Mihos, and Robert Dunbar for their different perspectives which enriched this final product.  I want to thank Kris Davidson and Roberta Humphreys for providing me with the time, environment, and encouragement to finish distilling this research into a publishable form.  This work was supported in part by the Jason J. Nassau Scholarship Fund and the Townsend Fund through the generous continued support of the Ford, Nassau, and Townsend families.

\appendix

\section{Appendix A:  Proto-planetary Nebula BD +33 2642}

BD +33 2642 is a known halo proto-planetary nebula (PPN) \citep{napiwotzki93,napiwotzki94} which has a spectrum with both classic nebular emission lines and a stellar photospheric absorption lines.  This gives the unique opportunity to analyze both the nebula and the central star.  Both the nebula and the central star are thought to be significantly iron deficient with enhancements of carbon, nitrogen, oxygen, and $\alpha$-process elements \citep{napiwotzki94,napiwotzki01}.  The abundances derived for the central star by this study and by \citet{napiwotzki94} for the central star and nebula are listed in Table \ref{tab13}.  

\begin{figure}
\figurenum{13}
\label{fig14}
\plotone{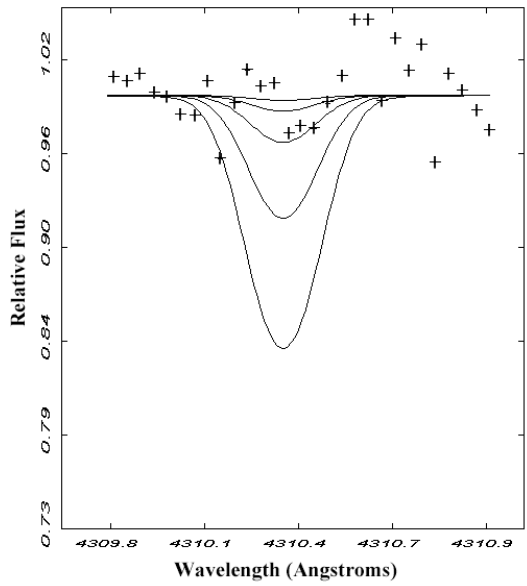}
\caption{Fitting the Fe III 4310 Å line.  The profiles from deepest to shallowest are: $log(\epsilon_{Fe})$= 7.50, 7.00, 6.50, 6.00, 5.50.  The $log(\epsilon_{Fe})$= 6.5 profile appears to be the upper limit on the iron abundance.}
\end{figure}

\citet{napiwotzki01} reassessed the iron abundance of the central star using a ultraviolet spectra from the HST-STIS spectrograph and found a slightly higher abundance of -1.5 dex relative to solar.  A higher iron abundance is not ruled out by the observed nebular lines since they can also be reproduced from model with [Z/H] = -1.3 \citep{napiwotzki94}.  The data at hand can be used to set an upper limit on the iron abundance using the \ion{Fe}{3} 4310 Å line.   Figure \ref{fig14} shows that the iron abundance can be no higher than about 6.50 dex ([Fe/H]=-1.0).  

Both the carbon (from \ion{C}{2}) and silicon (from \ion{Si}{3}) abundances derived by this study are higher than those reported by \citet{napiwotzki94}.   However the abundance measured from \ion{Si}{3} is more sensitive to the parameters adopted for the atmosphere.  Table \ref{tab14} compares the abundances derived from the equivalent widths in this study using the atmospheric parameters adopted for BD +33 2642 by \citet{napiwotzki94} and \citet{fitzmassa99}.  The abundances derived from \ion{C}{2}, \ion{N}{2}, and \ion{Al}{3} are fairly insensitive to the choice of model parameters while abundances derived from \ion{O}{2}, \ion{Si}{3}, \ion{S}{3}, and \ion{Ar}{2} can be significantly affected.  When the model parameters of \citet{napiwotzki94} are used, the silicon abundance as derived from \ion{Si}{3} is in excellent agreement with the value which they published.  However, the carbon abundance from \ion{C}{2} remains higher.

The difference in carbon abundance is better explained by differences in atomic data.  \citet{napiwotzki94} attempted to use the \ion{C}{2} $4267 \mbox{\AA}$ feature as one of four lines to measure the carbon abundance.  This feature cannot be used to reliably measure LTE carbon abundance \citep{kane80}.  Two of the other three features (\ion{C}{2} $6578 \mbox{\AA}$ and \ion{C}{2} $6583 \mbox{\AA}$) used by \citet{napiwotzki94} are also used by this study.  However, they use gf values which are significantly larger (Table \ref{tab8}).  Therefore the difference in the carbon abundance is a result of the line choice and atomic data. 

The differences in the carbon and silicon abundance have an impact on the interpretation of BD +33 2642.  The higher carbon abundance translates into a significant carbon enhancement relative to nitrogen and oxygen which requires a reassessment of the nuclear processing and mixing that may have occurred in the star.  

Likewise a higher silicon abundance may require a change in the grain formation scenario which is invoked to explain the lowered iron and aluminum abundances in the central star.  Because different processes affect grain formation for different elements, it is not expected that silicon will be lowered to the same degree as iron or aluminum.  However, there is no evidence from the silicon abundance that grain formation occurred since the silicon abundance is not lowered significantly with respect to the non-refractory elements argon and sulfur.  It is not clear if any grain formation scenario could lower the abundance of aluminum and iron and leave silicon untouched. 

\citet{napiwotzki01} also reported evidence for long-term (on the order of many years) variability in the radial velocity of the central star, which they cite as evidence of an unseen companion.  However, the radial velocity data in this study does not conclusively show any variability.  Table \ref{tab15} shows that there is a $4.8 \pm 10.6$ km/s difference in the radial velocity measured for BD +33 2642 in July 1998 and March 2001.  The heliocentric radial velocity measured in July 1998 ($-92.3\pm2.1$ km/s) is consistent with the systemic velocity adopted by \citet{napiwotzki01} (-94 km/s).  There is enough uncertainty in this measurement that a binary companion cannot be completely ruled out however this data does appear to conflict with the the model presented by \citet{napiwotzki01}.




\clearpage


\clearpage

\begin{deluxetable}{lrlrrrrrrr}
\tablewidth{0pt}
\tabletypesize{\tiny}
\tablecaption{Sample of Study\label{tab1}}
\tablecolumns{10}
\tablehead{
\colhead{Hipparcos}&\colhead{FB\tablenotemark{a}}&\colhead{Common}&\colhead{RA}&\colhead{Dec}&\colhead{Spectral}&\colhead{}&\colhead{Z\tablenotemark{b}}&\colhead{Gal Lat}&\colhead{Parallax}\\
\colhead{Number}&\colhead{Number}&\colhead{Designation}&\colhead{(J2000)}&\colhead{(J2000)}&\colhead{Class}&\colhead{V mag}&\colhead{(kpc)}&\colhead{(deg)}&\colhead{(mas)}}
\startdata
1214& 5& HD 1112& 00 15 27.28& -03 39 15.1& B9& 9.110& 0.43& -44.33&$1.67\pm1.22$\\
1511& \nodata& \nodata& 00 18 49.79& +30 23 33.7& B8& 11.610& 1.22& -31.95&$0.37\pm2.82$\\
3013& 7& BD -15 115& 00 38 20.26& -14 59 54.2& B2& 10.879& 4.42& -77.49&$1.52\pm1.80$\\
6419& 16& HD 8323& 01 22 24.88& -07 17 13.0& A0& 9.610& 0.59& -68.84&$1.19\pm1.31$\\
6727& 17& BD -7 230& 01 26 38.44& -06 20 18.7& \nodata& 11.130& 1.54& -67.59&$4.64\pm2.22$\\
11809& 22& Feige 23& 02 32 17.19& +02 59 52.7& B6V& 11.080& 1.94& -51.28&$1.68\pm4.25$\\
11844& 23& HD 15910& 02 32 54.48& -20 20 12.0& B9IV/V& 10.105& 1.05& -65.85&$0.81\pm1.64$\\
12320& 25& \nodata&02 38 37.78& +05 28 11.3& B7& 12.040& 2.52& -48.36&$3.92\pm5.81$\\
\enddata
\tablenotetext{a}{\citet{gs74}}
\tablenotetext{b}{Based on main-sequence spectroscopic parallax.}
\end{deluxetable}

\begin{deluxetable}{lrrrrr}
\tablewidth{0pt}
\tablecolumns{2}
\tablecaption{Abundance Control Samples\label{tab2}}
\tablehead{
 \colhead{}&\colhead{RA}&\colhead{Dec}&\colhead{Spectral}&\colhead{}&\colhead{V sin(i)}\\
 \colhead{Star}&\colhead{(J2000)}&\colhead{(J2000)}&\colhead{Class}&\colhead{V mag}&\colhead{(km/s)}}
\startdata
\cutinhead{Population I B Star Control Sample}\\
Gam Peg& 00 13 14& +15 11 01& B2 IV& 2.83& 3\\
HR 1220& 03 57 51& +40 00 37& B0.5 V& 2.90& 153\\
HR 1397& 04 26 21& +08 35 25& B6 IV& 6.07& 25\\
Bet Ori& 05 14 32& -08 12 06& B8Iab & 0.12& 33\\
Gam Ori& 05 25 08& +06 20 59& B2 III& 1.64& 59\\
HR 2010& 05 49 33& +12 39 05& B9 IV& 4.88& 22\\
HR 2928& 07 36 41& -19 42 08& B2 II& 5.69& 38\\
HR 4187& 10 40 35& +57 27 44& A0 V& 5.79& 27\\
HR 6165& 16 35 53& -28 12 58& B0 V& 2.81& 24\\
Nu Cap & 20 20 40& -12 45 33& B9 IV& 4.75& 17\\
HR 8404& 22 03 19& +11 23 12& B9.5 V& 5.82& 2\\
HR 8768& 23 02 45& +44 03 32& B2 V& 6.41& 25\\
\cutinhead{BHB Control Sample}\\
HD 60778& 07 36 12& -00 08 16& A1 V& 9.12& 10\\
HD 74721& 08 45 59& +13 15 49& A0 V& 8.71& 3\\
\cutinhead{PAGB Control Sample}\\
HD 18100& 02 53 41& -26 09 20& B5II/III& 8.44& 220\\
HD 44179& 06 19 58& -10 38 15& B8 V& 9.02& 20\\
HD 105262& 12 07 11& +12 59 08& B9& 7.09& 6\\
HD 119608& 13 44 31& -17 56 13& B1Ib & 7.50& 50\\
\enddata
\end{deluxetable}

\begin{deluxetable}{lrrrrrrr}
\tablewidth{0pt}
\tabletypesize{\tiny}
\tablecolumns{8}
\tablecaption{Results From Temperature \& Gravity Relations\label{tab3}}
\tablehead{
&\multicolumn{2}{c}{Extinction}&\colhead{Johnson\tablenotemark{2}}&
\colhead{Str\"{o}mgren\tablenotemark{3}}&\multicolumn{2}{c}{Geneva\tablenotemark{4}}&
\colhead{Balmer}\\
\colhead{Star}&\colhead{E(B-V)}&\colhead{Ref\tablenotemark{1}}&\colhead{$T_{eff}$}&
\colhead{$T_{eff}$}&\colhead{$T_{eff}$}&\colhead{log(g)}&\colhead{log(g)}
}
\startdata
\multicolumn{8}{c}{Program Stars}\\
\cline{1-8}\\
HD 1112& 0.038& b& 5703& & $11455\pm51$&$4.15\pm0.09$& 4.00\\
HIP 1511& 0.065& a& 7600& 9090& & & 3.00\\
BD -15 115& 0.018& a& 20322& 19427& $19758\pm198$& $3.94\pm0.17$& 4.00\\
HD 8323& 0.043& a& 12906& & $13568\pm80$& $4.05\pm0.13$\\
BD -07 230& 0.036& a& 9876& & & & 3.75\\
Feige 23& 0.028& a& 5728& 13265& & & 4.25\\
\enddata
\tablenotetext{1}{Sources for extinction are:  a=\citet{schlegel98}, b=\citet{kilkenny80}, c=\citet{savage85}, d=\citet{keenan83}, e=\citet{fitzpatrick99}, f=\citet{winkler97}, g=\citet{zorec91}, h=\citet{vogt98}, i=\citet{smalley95}, j=\citet{cardelli91}, k=\cite{eggen84}, l=\citet{eggen98}, m=\citet{huenemoerder84}, n=\citet{gray96}, o=\citet{pettini82},p=\citet{feltz72}}
\tablenotetext{2}{\citet{nsw93} (B-V) relation}
\tablenotetext{3}{\citet{nsw93} [u-b] relation.  The statistical error of the temperature determination (for one star) is approximately 3\%.}
\tablenotetext{4}{\citet{kunzli97} relation for stars with [Fe/H]=0.0}
\tablenotetext{5}{A nonsensical surface gravity value was produced that is probably indicative of a low gravity value for the star.}
\end{deluxetable}

\begin{deluxetable}{lrrrlr}
\tablewidth{0pt}
\tabletypesize{\tiny}
\tablecolumns{6}
\tablecaption{Stellar Parameters Used in the Analysis\label{tab4}}
\tablehead{
\colhead{}&\colhead{$T_{eff}$}&\colhead{Log(g)}&\multicolumn{2}{c}{Microturb ($\xi_{t}$)}&\colhead{Vsin(i)}\\ \colhead{Star}&\colhead{(K)}&\colhead{(dex)}&\colhead{(km/s)}&\colhead{Species\tablenotemark{b}}&\colhead{(km/s)}}
\startdata
\cutinhead{Program Stars}\\
HD 1112& 11500& 4.00& 10& Fe II&130\\
HIP 1511& 9000& 3.00& 2& Fe II, Cr II&20\\
BD -15 115& 19750& 4.00& 8& Si III&10\\
BD -07 230& 10000& 3.00& 2& Fe II, Cr II, Ti II&6\\
Feige 23& 13250& 4.25& 5& borrowed&240\\
HD 15910& 11250& 4.00& 5& Fe II&130\\
HIP 12320& 13500& 4.00& 5& borrowed&225\\
\enddata
\tablenotetext{a}{HD 138503 us an eclipsing binary which due to its unique nature, was analyzed in a slightly different manner.  For details see \citet{itlib}.}
\tablenotetext{b}{Species used to determine $\xi_{t}$.  Si III denotes the use of the Si III $4450\mbox{\AA}$ triplet rather than all the available lines of that species.}
\end{deluxetable}

\begin{deluxetable}{rrrrrrrr}
\tablewidth{0pt}
\tabletypesize{\tiny}
\tablecolumns{8}
\tablecaption{Abundance Changes as an Effect of Model Choice\label{tab5}}
\tablehead{\multicolumn{8}{c}{HR 6165}}
\startdata
&&\multicolumn{6}{c}{Log(g)}\\
&&\multicolumn{2}{c}{3.50}&\multicolumn{2}{c}{3.75}&\multicolumn{2}{c}{4.00}\\
\cline{3-8}\\
&&0.25&0.35&-0.03&0.20&0.07&0.10\\
&31500&0.43&0.70&0.19&0.35&0.05&0.14\\
&&0.33&0.58&0.14&0.31&0.03&0.15\\
&&0.20&0.48&0.07&0.24&0.01&0.10\\
\cline{3-8}\\
\enddata
\tablenotetext{*}{The center box in each chart lists the elements tested.  The other boxes list the differential abundance with respect to the center grid point for each element in the same order they are listed in the center box.}
\end{deluxetable}

\begin{deluxetable}{lrrrrrrrrrr}
\tablewidth{0pt}
\tablecolumns{11}
\tablecaption{Abundances of the Population I Control Sample\label{tab6}}
\tabletypesize{\tiny}
\tablehead{
&\colhead{C}&\colhead{N}&\colhead{O}&\colhead{Ne}&\colhead{Mg}&\colhead{Al}
&\colhead{Si}&\colhead{S}&\colhead{Ar}&\colhead{Fe}
}
\startdata
\cutinhead{High Galactic Latitude B Stars}\\
Gam Peg&8.13&7.55&8.33&8.16&7.53&5.96&7.19&7.12&6.66&7.21\\
&$\pm$0.16&$\pm$0.10&$\pm$0.23&&$\pm$0.07&$\pm$0.02&$\pm$0.22&$\pm$0.21&$\pm$0.25&$\pm$0.29\\
HR 1220&\nodata&8.53&8.69&\nodata&\nodata&6.93&7.57&\nodata&\nodata&8.02\\
&&$\pm$0.29&$\pm$0.23&&&&$\pm$0.06\\
HR 1397&8.73&8.32&\nodata&8.41&7.29&6.53&7.40&7.22&6.70&7.08\\
&$\pm$0.23&$\pm$0.43&&$\pm$0.24&$\pm$0.09&&$\pm$0.16&$\pm$0.19&$\pm$0.01&$\pm$0.18\\
Bet Ori&8.17&7.74&8.30&8.31&7.53&6.04&7.02&7.17&6.35&8.59\\
\enddata
\tablenotetext{1}{\citet{grevesse98}}
\tablenotetext{2}{\citet{vrancken00}}
\tablenotetext{3}{\citet{kilian92}}
\tablenotetext{4}{\citet{gl92}}
\tablenotetext{5}{\citet{daflon99}}
\tablenotetext{6}{\citet{daflon01}}
\end{deluxetable}

\begin{deluxetable}{lrr}
\tablewidth{0pt}
\tablecolumns{3}
\tablecaption{$T_{eff}$ Limits Imposed on Species in the Abundance Analysis\label{tab7}}
\tablehead{
\colhead{}&\colhead{Upper}&\colhead{Lower}\\
\colhead{Species}&\colhead{Limit (K)}&\colhead{Limit (K)}\\
}
\startdata
Ne II&\nodata&17000\\
Al II&20000&\nodata\\
Si III&\nodata&15000\\
Si II&\nodata&15000\\
S III&\nodata&13000\\
Ar II&\nodata&12000\\
Ti II&15000&\nodata\\
Cr II&15000&\nodata\\
Fe II&18000&\nodata\\
\enddata
\end{deluxetable}

\begin{deluxetable}{lrrrrrr}
\tablewidth{0pt}
\tabletypesize{\tiny}
\tablecolumns{7}
\tablecaption{Average Normal Star\label{tab8}}
\tablehead{
\colhead{Species}&\colhead{$\lambda_{0}(\mbox{\AA})$}&\colhead{$E_i$}&\colhead{log(gf)}&\colhead{$Avg_{\epsilon}$\tablenotemark{a}}&\colhead{$\sigma_{\epsilon}$\tablenotemark{a}}&N\tablenotemark{a}\\
&&\colhead{(eV)}&&\colhead{(dex)}&\colhead{(dex)}
}
\startdata
C II&4313.106&23.1161&-0.3730&8.39&0.02&2\\
C II&4318.606&23.1141&-0.4070&8.62&0.09&3\\
C II&4374.272&24.6535&0.6343&8.19&0.16&3\\
C II&4376.562&24.6600&0.3533&8.11&0.30&2\\
C II&4411.200&24.6019&0.5172&8.26&0.10&4\\
C II&4411.520&24.6033&0.6721&7.99&0.18&2\\
C II&4637.630&20.1500&-1.3401&8.54&\nodata&1\\
C II&5133.280&20.7040&-0.2100&8.30&0.47&2\\
\enddata
\tablenotetext{a}{Computed from the nearby Population I B star control sample.}
\end{deluxetable}

\begin{deluxetable}{lrrrrrrrrrrrrrrrrrl}
\tablewidth{0pt}
\tablecolumns{19}
\tablecaption{Relative Abundances\label{tab9}}
\rotate
\tabletypesize{\tiny}
\setlength{\tabcolsep}{0.03in}
\tablehead{
\colhead{}&\multicolumn{10}{c}{Relative Abundance\tablenotemark{b}}&
 \multicolumn{7}{c}{Abundance Ratios\tablenotemark{c}}\\
\colhead{Star}&\colhead{C}&\colhead{N}&
  \colhead{O}&\colhead{Mg}&\colhead{Al}&\colhead{Si}&\colhead{S}&
  \colhead{Cr}&\colhead{Ti}&\colhead{Fe}&\colhead{N/Al}&\colhead{O/Al}&\colhead{C/O}&\colhead{N/O}&
  \colhead{Si/Al}&\colhead{Mg/Fe}&\colhead{Si/S}&\colhead{Verdict\tablenotemark{a}}
}
\startdata
\cutinhead{High Galactic Latitude B Stars}\\
HD 1112&\nodata&\nodata&\nodata&\nodata&2.52&\nodata&\nodata&0.20&\nodata&-0.70&
\nodata&\nodata&\nodata&\nodata&\nodata&\nodata&\nodata&?\\
&&&&&&&&&&$\pm0.36$\\

HIP 1511&\nodata&\nodata&\nodata&\nodata&\nodata&\nodata&\nodata&-0.73&-1.25&-1.39&
\nodata&\nodata&\nodata&\nodata&\nodata&\nodata&\nodata&MP (-1)\\
&&&&&&&&$\pm0.24$&$\pm0.19$&$\pm0.34$\\

BD -15 115&-0.03&-0.24&0.13&-0.13&0.11&0.07&-0.28&\nodata&\nodata&0.12&
\em\underline{-0.35}\em&0.02&-0.16&-0.37&-0.04&-0.25&\em\underline{0.35}\em&Disk\\
&$\pm0.14$&$\pm0.21$&$\pm0.25$&$\pm0.02$&$\pm0.16$&$\pm0.21$&$\pm0.25$&&&
&$\pm$\em\underline{0.26}\em&$\pm0.30$&$\pm0.29$&$\pm0.33$&$\pm0.26$&$\pm0.02$&$\pm$\em\underline{0.33}\em\\

BD -07 230&\nodata&\nodata&\nodata&-0.27&\nodata&\nodata&\nodata&-0.13&-0.41&-0.39&
\nodata&\nodata&\nodata&\nodata&\nodata&0.12&\nodata&MP (-0.5)\\
\enddata
\tablenotetext{a}{See text for explanation of classifications.  ``Disk'' is a star consistent with the abundances in the nearby Pop I control sample.  ``MP'' notes a star that is metal poor with the measured abundances uniformly depressed.  ``OES'' is a star with abundance patterns consistent with older evolved stars.  Value in parenthesis next to metal poor (MP) classification is roughly the amount the abundances are depressed relative to the nearby Pop I control sample.  A question mark (?) denotes a level of uncertainty in the classification.}
\tablenotetext{b}{Abundances figured relative to the nearby Pop I control sample.  See section 4.6}
\tablenotetext{c}{Abundance ratios figured from relative abundances (See section 4.7).  The ratios which are significantly outside the range of values measured in the nearby Population I control sample (more than two standard deviations above or below the average) are underlined and in italics.  If the ratio was not measured the box is left blank.}
\tablenotetext{d}{See \citep{itlib}.}
\end{deluxetable}

\begin{deluxetable}{lrr}
\tablewidth{0pt}
\tablecolumns{3}
\tablecaption{Improvement of Abundance Errors by Relative Analysis\label{tab10}}
\tablehead{
\colhead{}&\colhead{Avg\tablenotemark{1}}\\
\colhead{Species}&\colhead{$\Delta\sigma$}&\colhead{$\sigma_{abs}$}
}
\startdata
C II&-0.02&0.29\\
N II&-0.01&0.22\\
O II&0.01&0.31\\
Mg II&0.02&0.13\\
Al III&-0.13&0.26\\
Si II&0.15&0.27\\
Si III&-0.03&0.23\\
S II&0.02&0.30\\
S III&-0.10&0.37\\
Ti II&-0.14&0.38\\
Cr II&-0.10&0.23\\
Fe II&-0.01&0.27\\
Fe III&-0.07&0.30\\
\enddata
\tablenotetext{1}{Avg $\Delta\sigma$ is the average value of ($\sigma_{abs} -  \sigma_{rel}$) for a given species.  A negative number means that the relative abundance analysis yielded a smaller standard deviation than the absolute abundance analysis.}
\end{deluxetable}

\begin{deluxetable}{llll}
\tablewidth{0pt}
\tablecolumns{4}
\tablecaption{Summary of Results\label{tab12}}
\tablehead{
&\colhead{Abundance}&&\colhead{Preliminary}\\
\colhead{Star}&\colhead{Class\tablenotemark{a}}&
 \colhead{Other\tablenotemark{b}}&\colhead{Classification\tablenotemark{c}}
}
\startdata
\cutinhead{Sample of Study}
HD 1112&?&&?\\
HIP 1511&MP (-1)&&?\\
BD -15 115&Disk&&Pop I\\
HD 8323&&&?\\
BD -07 230&MP (-.5)&&?\\
Feige 23&&NaD Em&?\\
HD 15910&MP (-.5)&&?\\
HIP 12320&Disk?&&Pop I\\
\enddata
\tablenotetext{a}{See Section 4.8.}
\tablenotetext{b}{Notes regarding any unusual spectral features.}
\tablenotetext{c}{Preliminary classification of the star based on the data presented in this paper.  OES = old evolved star.  Pop I = Population I runaway.  ? = unknown.}
\end{deluxetable}

\begin{deluxetable}{lcccccc}
\tablewidth{0pt}
\tablecolumns{7}
\tablecaption{Comparison of Abundances Derived for PPN BD +33 2642\label{tab13}}
\tablehead{
&\multicolumn{3}{c}{This Study}\\
&\multicolumn{3}{c}{Central Star\tablenotemark{c}}&
\multicolumn{2}{c}{NHK94\tablenotemark{a}}&\colhead{Nearby Pop I B Star}\\
&\colhead{Avg}&\colhead{$\sigma$}&\colhead{N}&\colhead{Central Star\tablenotemark{d}}&\colhead{Nebula}&\colhead{Control Sample}
}
\startdata
C&8.15&0.32&3 &$7.4 \pm 0.4$&8.5\tablenotemark{b}&$8.27 \pm 0.21$\\
N&7.47&0.19&13 &$7.3 \pm 0.4$&7.0\tablenotemark{b}&$7.99 \pm 0.34$\\
O&8.16&0.25&26 &$8.1 \pm 0.4$&\nodata&$8.55 \pm 0.16$\\
Al&5.41&0.41&3 &\nodata& \nodata&$6.31 \pm 0.33$\\
Si&7.07&0.11&3 &$6.7 \pm 0.3$&\nodata&$7.33 \pm 0.18$\\
S&6.28&0.12&2 &\nodata& \nodata&$6.88 \pm 0.20$\\
Ar&6.78&\nodata&1&\nodata& \nodata&$6.58 \pm 0.16$\\
Fe&$\leq6.5$\tablenotemark{*}&\nodata& \nodata&$5.5 \pm 0.5$&5.5\tablenotemark{b}&$7.46 \pm 0.30$\\
\enddata
\tablenotetext{a}{NHK94 = \citet{napiwotzki94}}
\tablenotetext{b}{The best fit model to the observed nebular spectra has C and N enhanced relative to [Z/H] = -2.0}
\tablenotetext{c}{Assumes parameters:  $T_{eff}=21500$ K, log(g)= 3.25 dex, $\xi_t=7$ km/s}
\tablenotetext{d}{Assumes parameters:  $T_{eff}=22000$ K, log(g)=2.9 dex, $\xi_t=10$ km/s}
\tablenotetext{*}{Upper limit (See Figure 12)}
\end{deluxetable}

\begin{deluxetable}{lrrr}
\tablewidth{0pt}
\tablecolumns{4}
\tablecaption{Effect of Model Parameters on Abundances of PPN BD +33 2642\label{tab14}}
\tablehead{
\colhead{}&\colhead{This Study}&\colhead{NHK94\tablenotemark{a}}&
\colhead{FM99\tablenotemark{b}}
}
\startdata
$T_{eff}$ (K)&21500&22000&18250\\
log(g)(dex)&3.25&3.00&2.50\\
$\xi_t$ (km/s)&7.0&10.0&6.3\\
C II&8.15&8.11&8.12\\
N II&7.47&7.43&7.55\\
O II&8.16&7.96&8.45\\
Al III&5.42&5.44&5.38\\
Si III&7.07&6.70&7.37\\
S III&6.28&6.11&6.38\\
Ar II&6.78&6.97&6.62\\
\enddata
\tablenotetext{a}{NHK94 = \citet{napiwotzki94}}
\tablenotetext{b}{FM99 = \citet{fitzmassa99}}
\end{deluxetable}

\begin{deluxetable}{lrrr}
\tablewidth{0pt}
\tablecolumns{4}
\tablecaption{Relative Velocity for BD +33 2642: July 1998 vs March 2001\label{tab15}}
\tablehead{
&\colhead{July 1998}&\colhead{$\Delta\lambda$\tablenotemark{b}}&
\colhead{$\Delta V$\tablenotemark{b}}\\
\colhead{Species}&\colhead{Rest $\lambda$\tablenotemark{a}}&
\colhead{($m\mbox{\AA}$)}&\colhead{(km/s)}
}
\startdata
S II&$5639.977 \mbox{\AA}$&-134&-7.1\\
S II&$5660.001\mbox{\AA}$&-259&-13.7\\
N II&$5679.558\mbox{\AA}$&153&+8.1\\
N II&$5679.558\mbox{\AA}$&-5&-0.3\\
N II&$5686.213\mbox{\AA}$&-415&-21.9\\
Al III&$5696.604\mbox{\AA}$&-82&-4.3\\
Al III&$5722.730\mbox{\AA}$&-103&-5.4\\
Al III&$5722.730\mbox{\AA}$&-184&-9.6\\
Si III&$5739.730\mbox{\AA}$&204&+10.7\\
C II&$6578.050\mbox{\AA}$&312&+14.2\\
N II&$6610.562\mbox{\AA}$&-34&-15.4\\
N II&$6610.562\mbox{\AA}$&-444&-20.1\\
He I&$6678.151\mbox{\AA}$&-92&-4.1\\
O II&$6721.388\mbox{\AA}$&39&1.7\\
\cline{1-4}\\
\multicolumn{2}{r}{Average}&-96&$-4.8\pm10.6$\\
\enddata
\tablenotetext{a}{Assumes a heliocentric radial velocity of $-92.3\pm2.1$ km/s}
\tablenotetext{b}{Difference between observations on Julian Dates 2451013.6716 and 2451950.9534}
\end{deluxetable}

\end{document}